\documentclass[%
 aip,
 amsmath,amssymb,
 reprint,%
]{revtex4-1}

\usepackage{graphicx}
\usepackage{dcolumn}
\usepackage{bm}
\usepackage{float}
\usepackage{lineno}
\usepackage{graphicx}
\usepackage{xcolor}
\usepackage[utf8]{inputenc}
\usepackage[T1]{fontenc}
\usepackage{mathptmx}
\usepackage{etoolbox}
\makeatletter
\def\@email#1#2{%
 \endgroup
 \patchcmd{\titleblock@produce}
  {\frontmatter@RRAPformat}
  {\frontmatter@RRAPformat{\produce@RRAP{*#1\href{mailto:#2}{#2}}}\frontmatter@RRAPformat}
  {}{}
}%
\makeatother
\begin{document}

\preprint{AIP/123-QED}

\title{Investigation of neutron imaging applications using fine-grained nuclear emulsion}

\author{Abdul Muneem}
\affiliation{High Energy Nuclear Physics Laboratory, Cluster for Pioneering Research, RIKEN, 2-1 Hirosawa, Wako, 351-0198, Saitama, Japan}
\affiliation{GIK Institute of Engineering Sciences and Technology, Topi, Swabi, 23640, KP, Pakistan}
\author{Junya Yoshida}%
\affiliation{High Energy Nuclear Physics Laboratory, Cluster for Pioneering Research, RIKEN, 2-1 Hirosawa, Wako, 351-0198, Saitama, Japan}
\affiliation{Department of Physics, Tohoku University, Aramaki, Aoba-ku, 980-8578, Sendai, Japan}
\affiliation{International Center for Synchrotron Radiation Innovation Smart, Tohoku University, Sendai, 980-8577, Japan}
\author{Hiroyuki Ekawa}
\affiliation{High Energy Nuclear Physics Laboratory, Cluster for Pioneering Research, RIKEN, 2-1 Hirosawa, Wako, 351-0198, Saitama, Japan}
\author{Masahiro Hino}
\affiliation{Institute for Integrated Radiation and Nuclear Science, Kyoto University, Kumatroi, Sennan-gun, 590-0494, Osaka, Japan}
\author{Katsuya Hirota}
\affiliation{High Energy Accelerator Research Organization (KEK), Tsukuba, 305-0801, Ibaraki, Japan}
\author{Go Ichikawa}
\affiliation{High Energy Accelerator Research Organization (KEK), Tsukuba, 305-0801, Ibaraki, Japan}
\affiliation{Japan Proton Accelerator Research Complex (J-PARC) Center, Tokai, 319-1195, Ibaraki, Japan}
\author{Ayumi Kasagi}
\affiliation{High Energy Nuclear Physics Laboratory, Cluster for Pioneering Research, RIKEN, 2-1 Hirosawa, Wako, 351-0198, Saitama, Japan}
\affiliation{Graduate School of Engineering, Gifu University, 501-1193, Gifu, Japan}
\author{Masaaki Kitaguchi}           
\affiliation{Department of Physics, Nagoya University, Furo-cho, Chikusa, 464-8601, Nagoya, Japan}
\affiliation{Kobayashi-Maskawa Institute for Origin of Particles and the Universe(KMI), Nagoya University, Furo-cho, Chikusa, 464-8601, Nagoya, Japan} 
\author{Naoto Muto}
\affiliation{Department of Physics, Nagoya University, Furo-cho, Chikusa, 464-8601, Nagoya, Japan}
\author{Kenji Mishima}
\affiliation{High Energy Accelerator Research Organization (KEK), Tsukuba, 305-0801, Ibaraki, Japan}
\affiliation{Japan Proton Accelerator Research Complex (J-PARC) Center, Tokai, 319-1195, Ibaraki, Japan}
\author{Jameel-Un Nabi}
\affiliation{GIK Institute of Engineering Sciences and Technology, Topi, Swabi, 23640, KP, Pakistan}
\affiliation{University of Wah, Quaid Avenue, Wah Cantt, 47040, Punjab, Pakistan}
\author{Manami Nakagawa}
\affiliation{High Energy Nuclear Physics Laboratory, Cluster for Pioneering Research, RIKEN, 2-1 Hirosawa, Wako, 351-0198, Saitama, Japan}
\author{Naotaka Naganawa}
\affiliation{Department of Physics, Nagoya University, Furo-cho, Chikusa, 464-8601, Nagoya, Japan}
\affiliation{Institute of Materials and Systems for Sustainability, Nagoya University,
Furo-cho, Chikusa-ku, Nagoya, 464-8601, Japan}
\author{Takehiko R. Saito}
\affiliation{High Energy Nuclear Physics Laboratory, Cluster for Pioneering Research, RIKEN, 2-1 Hirosawa, Wako, 351-0198, Saitama, Japan}
\affiliation{GSI Helmholtz Centre for Heavy Ion Research, Planckstrasse 1, 64291, Darmstadt, Germany}
\affiliation{School of Nuclear Science and Technology, Lanzhou University, 222 South Tianshui Road, Lanzhou, 730000, Gansu Province, China}
\email{takehiko.saito@riken.jp}
\email{abdul.muneem@riken.jp}
\date{\today}
\begin{abstract}
Neutron imaging is a non-destructive inspection technique with a wide range of applications.  One of the  important aspects concerning neutron imaging is achieving micrometer-scale spatial resolution. Developing a neutron detector with a high resolution is a challenging task. Neutron detectors, based on fine-grained nuclear emulsion, may be suitable for  high resolution neutron imaging applications. High track density is a necessary requirement to improve the quality of neutron imaging. However, the available track analysis methods are difficult to apply under high track density conditions. Simulated images were used to determine the required track density for neutron imaging. It was concluded that a track density of the order of $10^4$ tracks per 100 $\times$ 100 $\mu$m$^2$ is sufficient to utilize neutron detectors for imaging applications. The contrast resolution was also investigated for the image data sets with various track densities and neutron transmission rates. Moreover, experiments were performed for neutron imaging of the gadolinium-based gratings with known geometries. The structure of  gratings was successfully resolved. The calculated 1$\sigma$ 10-90 \% edge response, using the gray scale optical images of the grating slit with a periodic structure of 9 $\mu$m, was 0.945 $\pm$ 0.004 $\mu$m. 
\end{abstract}

\maketitle

\section{\label{introduction}Introduction}
Neutron imaging (NI) is a non-destructive  technique for visualizing and analysing the inner structure of  material objects. It has a wide range of applications \cite{strobl2009, hussey2015,lehmann2015,kardjilov2018,tengattini2020,Song2019,Lv2018}. The basic principle for NI is similar to that of  X-ray imaging. Neutrons are neutral particles and can interact directly with atomic nuclei. The absorption and scattering cross-sections of neutrons depend on the nuclides. The dependence of attenuation coefficients of neutrons is different from that of X-rays which is correlated to the atomic number \cite{Banhart2008}. Contrary to X-rays, neutrons exhibit high attenuation for light (e.g. hydrogen, carbon, boron and lithium) and heavy (e.g. cadmium and gadolinium) nuclei. Lower attenuation is observed for heavy nuclei including aluminum, silicon, titanium and lead.
These characteristics give neutron imaging an advantage over X-ray imaging. Neutron imaging may be employed as an effective method to  visualize distributions for a wide range of elements using the difference in their neutron attenuation coefficients \cite{strobl2009}.

The major challenges in developing NI techniques include improvements in spatial resolution \cite{tremsin2011,hussey201501, trtik2015, trtik2016, bingham2015,Morgano18,trtik2020}, detection efficiency, \cite{tremsin2011} and time resolution \cite{kardjilov2018, woracek19} of neutron detectors.
Recent developments have improved the spatial resolution of NI to a few micrometers. One of the common approaches in NI is to convert the shading field of the neutron beam into an image of visible light using scintillators containing converters such as $^6$Li or $^{157}$Gd \cite{trtik2020,Lehmann_2011,hussey2017,JIANG20211942}.
The luminescence of a neutron capture event was observed by a charge--coupled device (CCD) or a complementary metal-oxide-semiconductor (CMOS) image sensor via an optical system.
Using a technique based on this approach, the best resolution resolution currently reported is 2 $\mu$m \cite{hussey2017}.
This spatial resolution was achieved using a thin gadolinium oxysulfide scintillator, magnifying optics, a CMOS image sensor, and an event positioning method by detecting the center of the luminescence for each event. Isegawa \emph{et al.}\cite{jimaging7110232} used a Gd${_3}$Al${_2}$Ga${_3}$O${_{12}}$:Ce single-crystal scintillator for the neutron radiography and achieved a spatial resolution of 10.5 $\mu$m.

For micrometer-scale spatial resolution or even better, nuclear emulsion is one of the potential candidates for NI device. Nuclear emulsion is a photographic film that can record the three-dimensional trajectories of charged particles with submicrometer-scale spatial resolution. It is composed of silver halide (AgBr) crystals dispersed in gelatin. Nuclear emulsions were initially developed for particle and nuclear physics experiments to observe the tracks of charged particles. They were used to observe pions \cite{lattes1947}, hypernuclei \cite{Danysz1953, DAVIS20053}, charm particles \cite{niu1971} and double strangeness hypernuclei \cite{DANYSZ1963121}. The analysis speed of nuclear emulsions has increased remarkably with improvements in image processing techniques \cite{yoshimoto2017,saito2021}. These improvements enable the use of nuclear emulsion  in modern experiments, and led to the first direct detection of tau-neutrinos \cite{KODAMA2001218}, discovery of $\nu{_\mu} \rightarrow \nu{_\tau}$
oscillations in appearance mode \cite{PhysRevLett.115.121802}, hypernuclei \cite{saito2021,Hiyama2018,Ekawa2019,Hayakawa2021,Yoshimoto2021} and for several other applications including muon radiography \cite{tanaka2007, Morishima2017}.

Since the last decade, it has become possible to produce dedicated nuclear emulsions for specific purposes.  Controlling the size of the AgBr crystals and chemical components contributed in achieving optimal spatial resolution and sensitivity. Fine-grained nuclear emulsions (FGNEs) with AgBr crystals, having a diameter of less than 50 nm, was developed \cite{naka2013, asada2017} to detect the tracks of the nuclei recoiled by the so-called weakly interacting massive particles. The sensitivity of the nuclear emulsion was optimised to track the nuclei but was maintained sufficiently low for electrons and $\gamma$-rays \cite{10.1093/ptep/ptab030}.

The FGNE with crystal diameter of 40 nm was employed for cold/ultracold neutron detection by combining it with a neutron converter layer formed by $^{10}$B${_4}$C. This layer absorbs neutrons and emits two charged particles heading in the opposite direction via the following process: $^{10}$B + n $\rightarrow$ $^{7}$Li + $^{4}$He. When one of the emitted ions passes isotropically through an emulsion layer, latent image specks are created in some AgBr$\cdot$I crystals along its trajectory. After the chemical development, each crystal that contains latent image specks changes to an enlarged silver grain. A series of these grains is recognized as a track of the charged particle. Naganawa \emph{et al.}\cite{naganawa2018} reported the development of a cold/ultracold neutron detector using FGNE and that the spatial resolution of neutron absorption points was estimated as less than 0.1 $\mu$m from linear fitting of tracks. The track lengths of $^{7}$Li and $^{4}$He recorded in the emulsion layer were 2.6 $\pm$ 0.4 and 5.1 $\pm$ 0.4 $\mu$m, respectively. This development was initiated for fundamental physics experiments to investigate quantized states under the influence of the earth’s gravitational field and to search unknown interactions attracting neutrons \cite{Nesvizhevsky2002, ABELE2009593c,Ichikawa2014,Muto_2022}.

The FGNEs combined with a boron-based neutron converter can be applied for NI applications. For the purpose of NI, the accumulated track density must be several orders of magnitude higher than that used in ref. \cite{naganawa2018}. Hirota \emph{et al.} \cite{hirota2021} demonstrated NI of a crystal oscillator chip using FGNE under the accumulated track density of $3 \times 10^4$ tracks per 100 $\times$ 100 $\mu$m$^2$. They used a similar NI detector as in ref. \cite{naganawa2018}, but the thickness of $^{10}$B${_4}$C layer was 2 $\mu$m to increase neutron conversion efficiency. The thickness of $^{10}$B${_4}$C layer used for neutron detection in ref. \cite{naganawa2018} was 50 nm. Gold wires with a diameter of approximately 30 $\mu$m in the crystal oscillator chip were successfully visualized. Furthermore, they attempted NI of a gadolinium-based grating with a periodic structure of 9 $\mu$m under the accumulated track density of $6 \times 10^3$ tracks per 100 $\times$ 100 $\mu$m$^2$. They confirmed the grating spacing of 9 $\mu$m pitch using Rayleigh test with a standard deviation of 1.3 $\mu$m. They concluded that the achievable resolution for the NI using the FGNE was better than 3 $\mu$m. The challenges presented in ref. \cite{hirota2021} were to develop a method of image analysis and  spatial resolution under a condition of $\sim$ $10^4$ tracks per 100 $\times$ 100 $\mu$m$^2$. With such a high track density, the conventional track-by-track analysis becomes difficult because of the overlap of the tracks.

For NI, the neutron detector based on the FGNE combined with neutron converter has several merits. It can detect neutron capture events with submicrometer-scale spatial resolution. The detector is lightweight, thin, small ( adjustable to sample objects with an area of approximately 100 cm$^2$), and does not require any electric supply or drive in electronics. Furthermore, this detector can be integrated into devices or sample environment apparatus to shorten the distance with the investigated object. It can be added to existing imaging systems without any major modification and may be used as a complementary imaging device. Along with these merits, there are several limitations as well. A longer beam exposure time is required for a thinner $^{10}$B${_4}$C layer in order to achieve higher spatial resolution. An additional process of chemical development is required after exposure. Furthermore, this detector is unsuitable for real-time measurements and applications where neutron energy measurement is required.

For NI applications using the FGNE, a high density accumulated tracks is a basic requirement. We describe our detector in section \ref{FGNE_detector}. The estimation of  track density and diffuseness as a metric to evaluate the imaging resolution is discussed in section \ref{width_resolution}. The procedure employed to investigate the contrast resolution is discussed in section \ref{estimation_contrast_resolution}.  The details of the experiments are discussed in section \ref{experiments}. The results for NI of gadolinium-based gratings are discussed in section \ref{results}. The NI application of the FGNE is discussed in the section \ref{Siemens_star}. We finally conclude our findings and discuss future prospects in 
section \ref{conclusion}.

\section{\label{FGNE_detector}Neutron detector}
\begin{figure}[htbp]
    \centering
    \includegraphics[width=\linewidth]{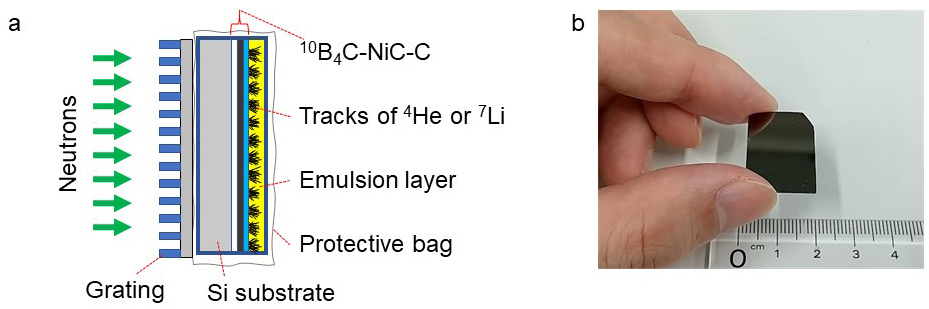}
    \caption{(a) Schematic layout of the neutron detector using a fine-grained emulsion packed in a protective bag combined with a gadolinium-based grating. (b) Photograph of the developed neutron  detector. The size of the detector is 25 mm $\times$ 15 mm.}
    \label{fig:emulsion_detector}
\end{figure}

The neutron detector developed for this work, shown in Figure \ref{fig:emulsion_detector}a,  has a layered structure. 
The base of the detector is a 0.4 mm thick silicon substrate. On this substrate, $^{10}$B${_4}$C, NiC and C layers of thicknesses 230 nm, 46 nm and 14 nm, respectively, were formed by ion beam sputtering technique \cite{hino2013}. The thickness of the $^{10}$B${_4}$C layer was determined by taking into account neutron detection efficiency, the physical stability, and the spatial resolution for the imaging.
The $^{10}$B${_4}$C layer converts an incident neutron into  $^{4}$He and $^{7}$Li. The NiC layer was used to physically stabilise the $^{10}$B${_4}$C layer. The C layer was used for chemical protection from the NiC layer and for providing a strong adhesion to the emulsion. An emulsion layer of thickness  10 $\mu$m was formed on the sputtered layers in a dark room by pouring and drying the FGNE. It was composed of AgBr$\cdot$I crystals of approximately 40 nm diameter uniformly dispersed in medium including gelatin and polyvinyl alcohol.  
The neutron detector was packed with a light- and air-tight laminated bag,  composed of nylon, polyethylene,  aluminum, polyethylene and black polyethylene layers having thicknesses of 15 $\mu$m, 13 $\mu$m, 7 $\mu$m, 13 $\mu$m and 35 $\mu$m, respectively. During the beam exposure, the detector and the gadolinium-based grating were placed close together to minimise the blurring effects induced by the spreading of transmitted neutron beam, as shown in Figure \ref{fig:emulsion_detector}a.
Some incident neutrons were absorbed in the grating, and remaining neutrons were absorbed in the $^{10}$B${_4}$C layer.
When a $^{10}$B nucleus absorbs a neutron, the reaction products, $^{4}$He and $^{7}$Li, are emitted head off in opposite directions. When one of them passes through the emulsion layer, latent image specks are created in  AgBr$\cdot$I crystals along its trajectory. After chemical development, crystals  containing latent image specks transform to silver grains.  Figure \ref{fig:emulsion_detector}b shows a photograph of the  detector after the chemical development. After chemical development, the thickness of the emulsion layer shrinks by a factor of 0.55 $\pm$ 0.1. 
A set of gray scale images were obtained with submicrometer precision using an epi-illumination optical microscope. The series of silver grains in these images is recognized as the induced tracks of the charged particles.

\section{Simulations for imaging applications}
Tracks of $^4$He
and $^7$Li have certain length as they travel through the nuclear
emulsion. If the track density is high ($\sim$ $10^5$ tracks per 100 $\times$ 100 $\mu$m$^2$), they blot out the edge of the object in the obtained images. On the other hand, if the track density is low ($\sim$ $10^2$ tracks per 100 $\times$ 100 $\mu$m$^2$), the outline of the object will not be visible. Therefore, it was important to estimate 
the required track density for NI, and diffuseness as a function of track density for the current NI system. The track density required for NI, and diffuseness was
estimated using simulated images which we present below.
\begin{figure*}[htbp]
    \centering
    \includegraphics[width=0.9\linewidth]{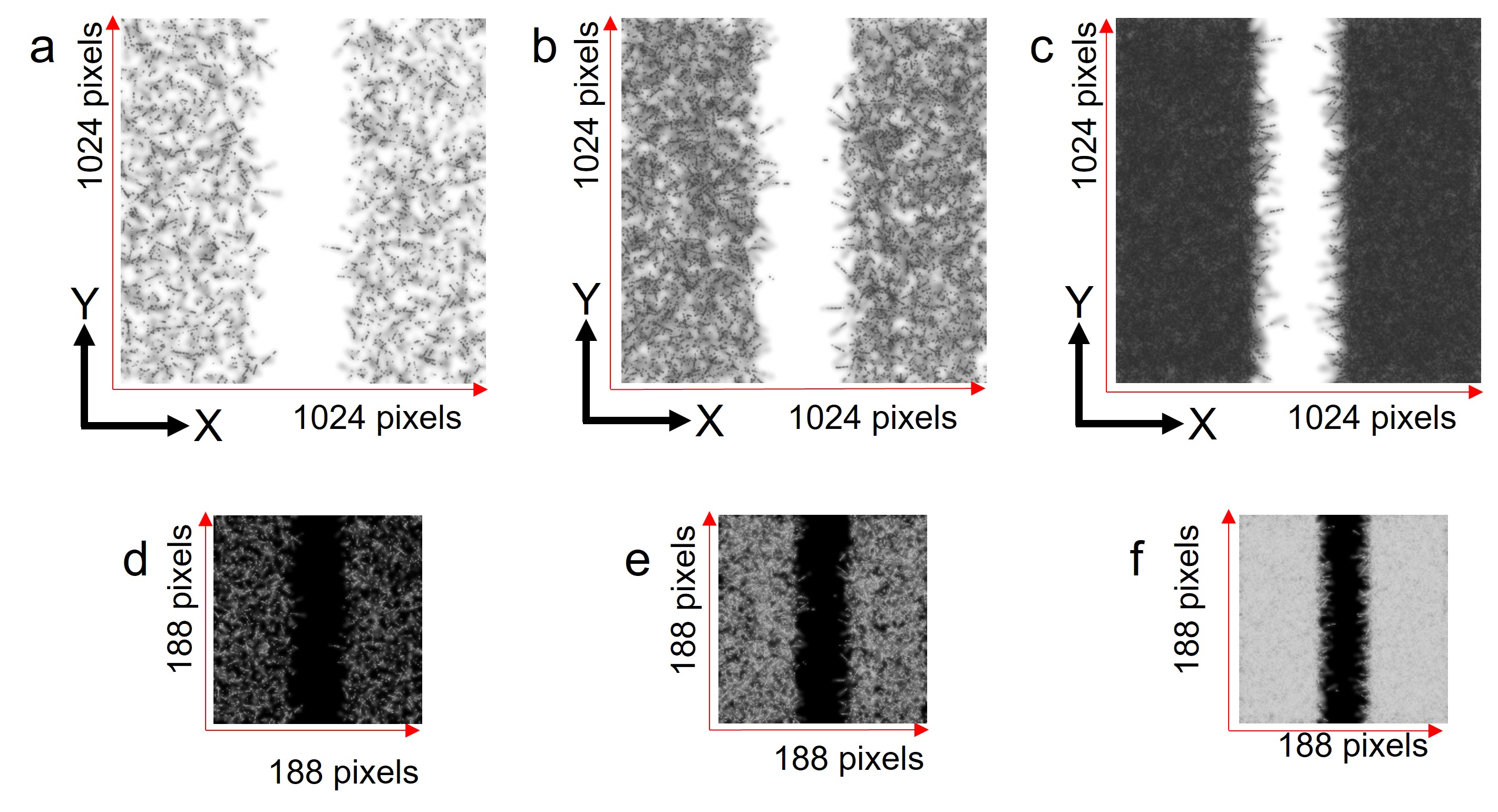}
    \caption{Simulated images for estimation of track density conditions for neutron imaging. (a) $1.0 \times 10^4$, (b) $3.0 \times 10^4$ and (c) $2.0 \times 10^5$, tracks per 100 $\times$ 100 $\mu$m$^2$. Images (d), (e) and (f) are the  inverted and resized images of (a), (b) and (c), respectively.}
    \label{fig:width_resolution_sim_imgs}
\end{figure*}

\begin{figure*}[htbp]
    \centering
    \includegraphics[width=0.9\linewidth]{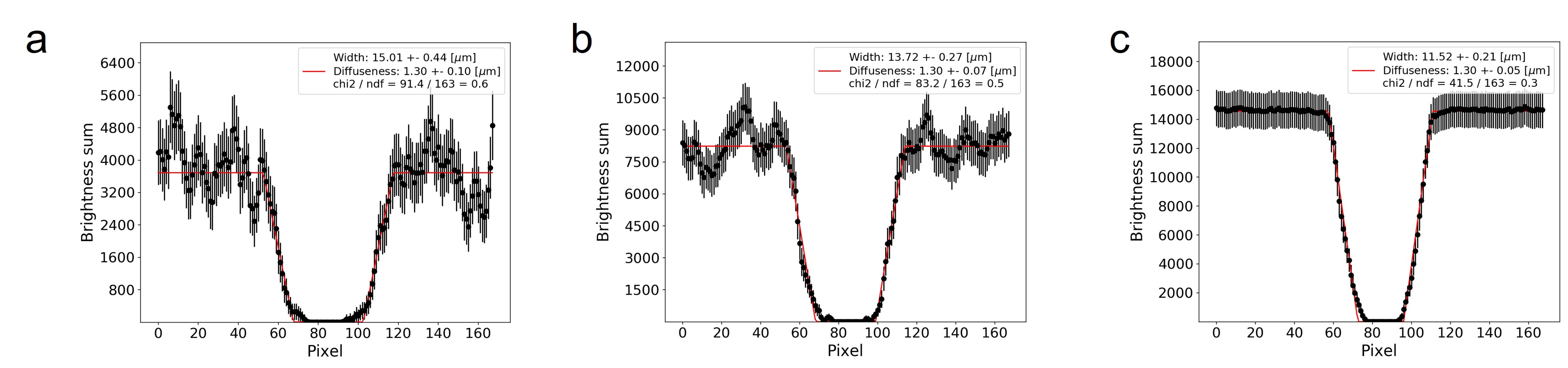}
    \caption{(a), (b) and (c) show the brightness sum in the simulated images shown in Figures \ref{fig:width_resolution_sim_imgs}d, \ref{fig:width_resolution_sim_imgs}e and \ref{fig:width_resolution_sim_imgs}f, respectively,  along the direction parallel to the grating (Y-direction) as a function of the X position, and trapezoid fitting for the single bar pattern with a width of 15 $\mu$m}
    \label{fig:width_resolution_trap_fit}
\end{figure*}

\begin{figure}[htbp]
    \centering
    \includegraphics[width=0.9\linewidth]{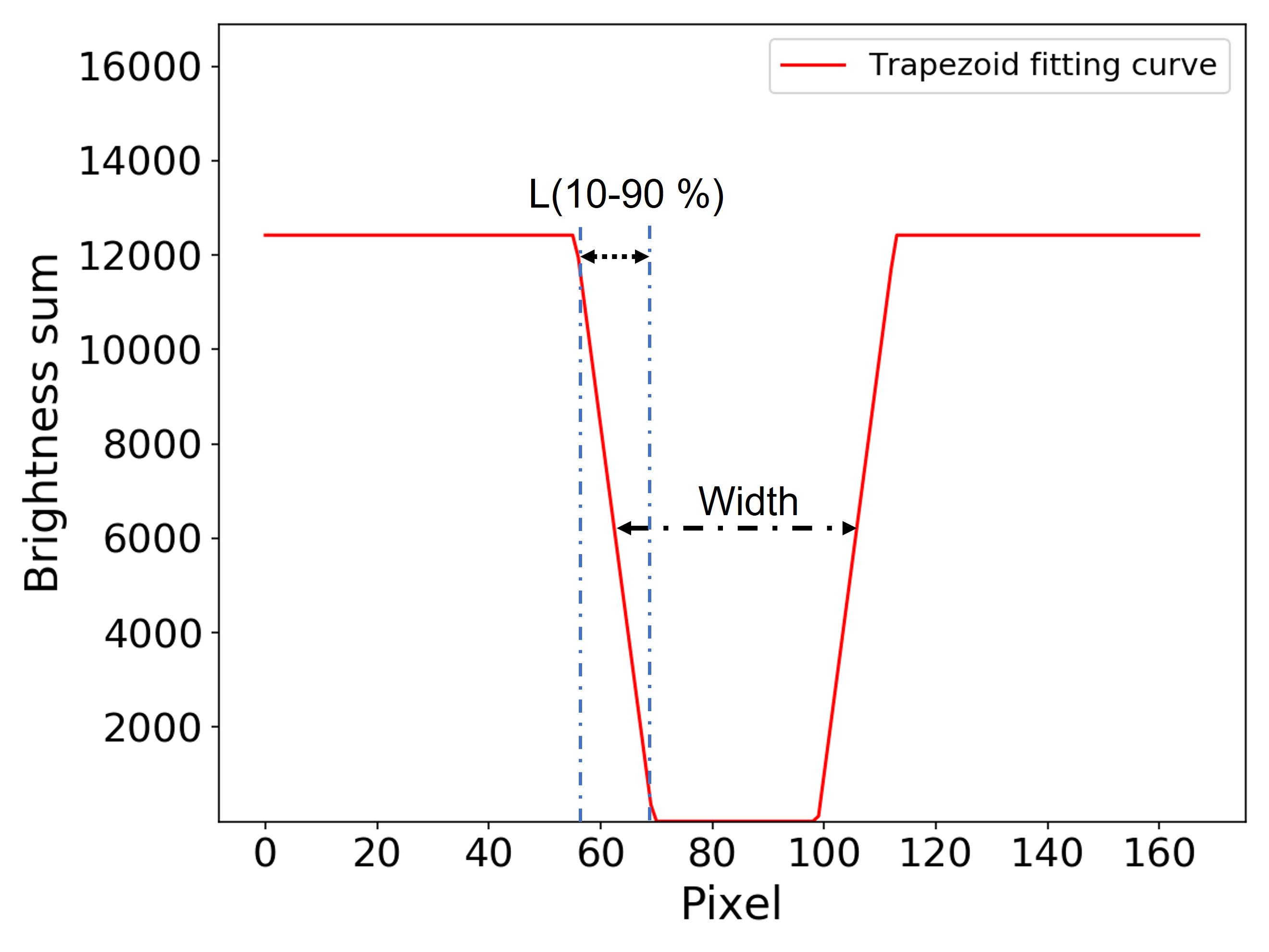}
    \caption{Example of a trapezoid curve used in this work. The width is the distance between the mid points midpoints of the slopes of trapezoid fitting curve. The 10-90 \% L(10-90\%) edge response is the distance between the 10 \% and 90 \% of the knife edges of the trapezoid fitting curve.}
    \label{fig:trap_fit_example}
\end{figure}

\subsection{\label{width_resolution}Estimation of track density}
Data sets of simulated images were generated under several track density conditions ranging from  $3 \times 10^3$ to $3 \times 10^5$ tracks per 100 $\times$ 100 $\mu$m$^2$. Each data set contained 165 images. The images are shown in Figures \ref{fig:width_resolution_sim_imgs}a, \ref{fig:width_resolution_sim_imgs}b and \ref{fig:width_resolution_sim_imgs}c for track densities of $1 \times 10^4$, $3 \times10^4$ and $2 \times 10^5$ tracks per 100 $\times$ 100 $\mu$m$^2$, respectively. Each simulated image had a single bar pattern with a width of 15 $\mu$m where the incidents neutrons are absorbed. Since the track length of $^{4}$He recorded in the emulsion layer was 5.1 $\pm$ 0.4 $\mu$m \cite{naganawa2018}, the width of single bar pattern was taken as 15 $\mu$m. As shown in the Figure \ref{fig:width_resolution_sim_imgs}c, decrease in the width of the bar pattern is visible which is due to finite length of the tracks with an increase in the track density. The simulated images were scaled down to 188 pixels $\times$ 188 pixels using the nearest neighbor method \cite{NearestNeighbor2012} where size of one pixel is equivalent to 0.3 $\mu$m (typical  diameter of a silver grain). Figures \ref{fig:width_resolution_sim_imgs}d, \ref{fig:width_resolution_sim_imgs}e and \ref{fig:width_resolution_sim_imgs}f show resized and inverted images of Figures \ref{fig:width_resolution_sim_imgs}a, \ref{fig:width_resolution_sim_imgs}b and \ref{fig:width_resolution_sim_imgs}c, respectively.
The trapezoid fitting for the projection of brightness sum of the corresponding resized and inverted images is shown in Figures \ref{fig:width_resolution_trap_fit}a, \ref{fig:width_resolution_trap_fit}b and \ref{fig:width_resolution_trap_fit}c, respectively. The width of single bar pattern was defined as the distance between the midpoints of the slopes of the trapezoid fitting curve. The edge response defined as L(10-90\%) was deduced as the distance between 10 \% and 90 \% of the knife edges of the trapezoid fitting curve as shown in Figure \ref{fig:trap_fit_example}.

\begin{figure}[htbp]
    \centering
    \includegraphics[width=\linewidth]{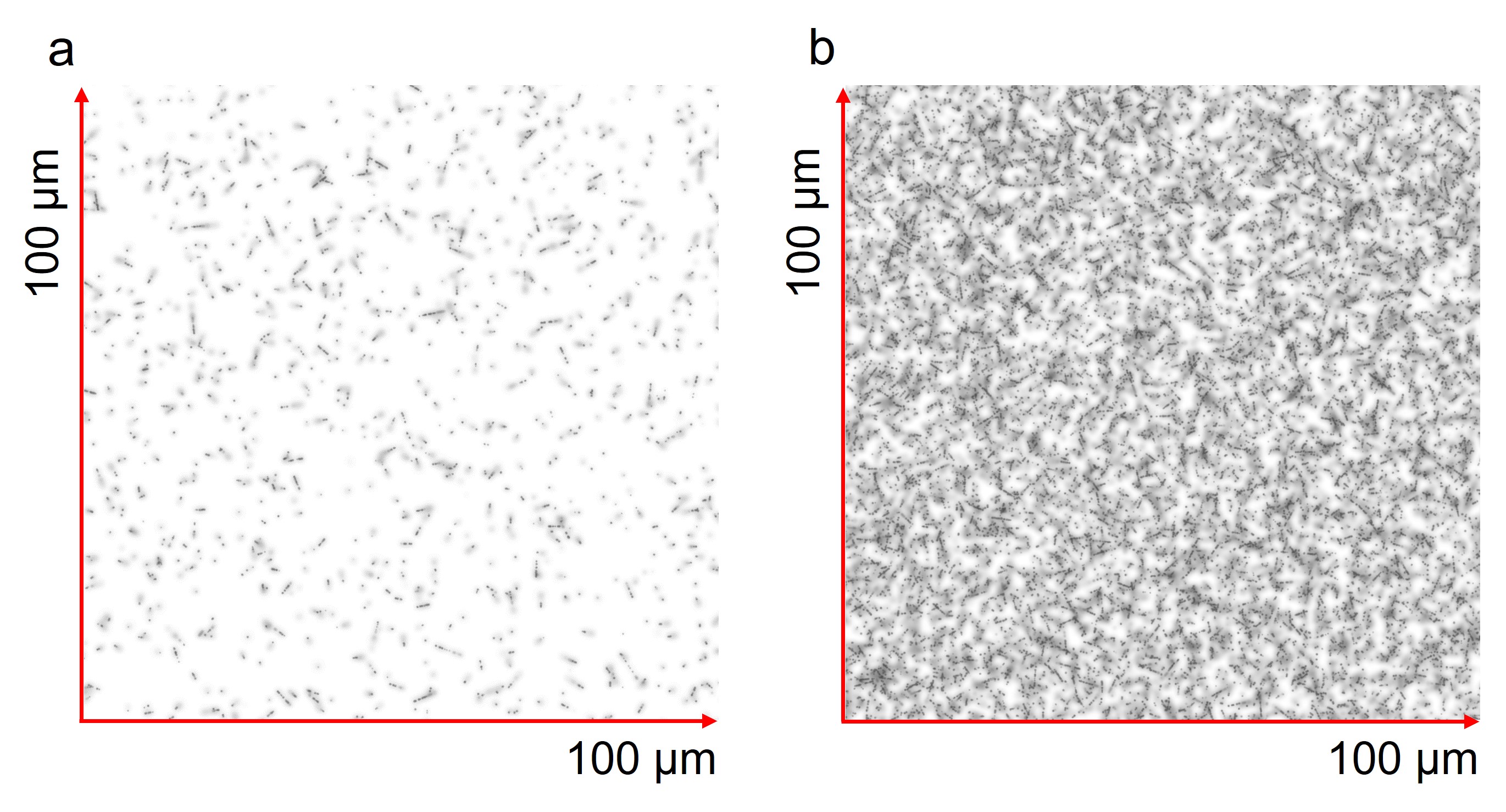}
    \caption{Simulated images for the estimation of the statistical error bars for the corresponding brightness values in the acquired images. The track densities are (a) $1 \times$ $10^3$, and (b) $1 \times$ $10^4$ per 100 $\times$ 100 $\mu$m$^2$ }
    \label{fig:errro_bar_ana_imgs}
\end{figure}

\begin{figure}[htbp]
    \centering
    \includegraphics[width=\linewidth]{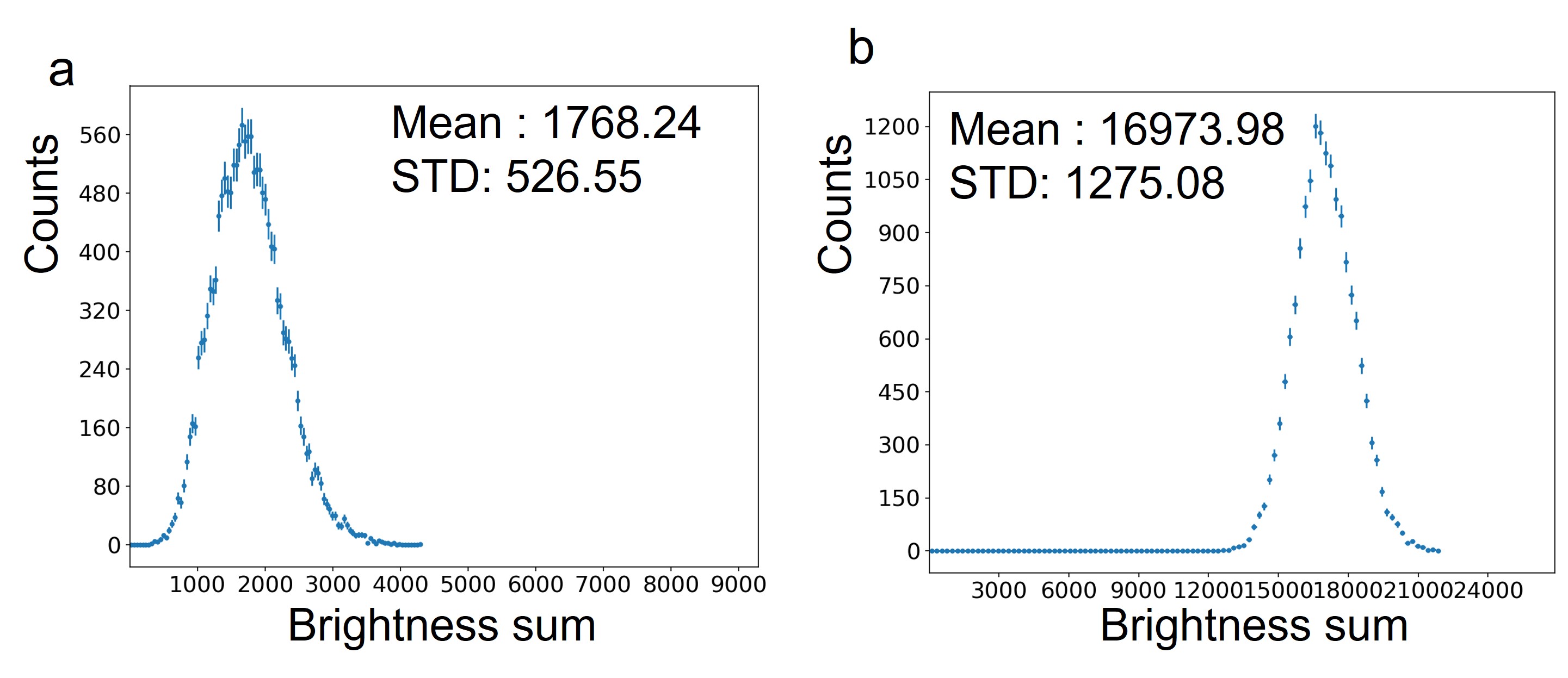}
    \caption{Projection of the brightness sum of the simulated images for the statistical error bar estimation of the corresponding brightness values  (a) as shown in Figure \ref{fig:errro_bar_ana_imgs}a, and (b) as shown in Figure \ref{fig:errro_bar_ana_imgs}b}
    \label{fig:errro_bar_ana_projection}
\end{figure}

\begin{figure}[htbp]
    \centering
    \includegraphics[width=\linewidth]{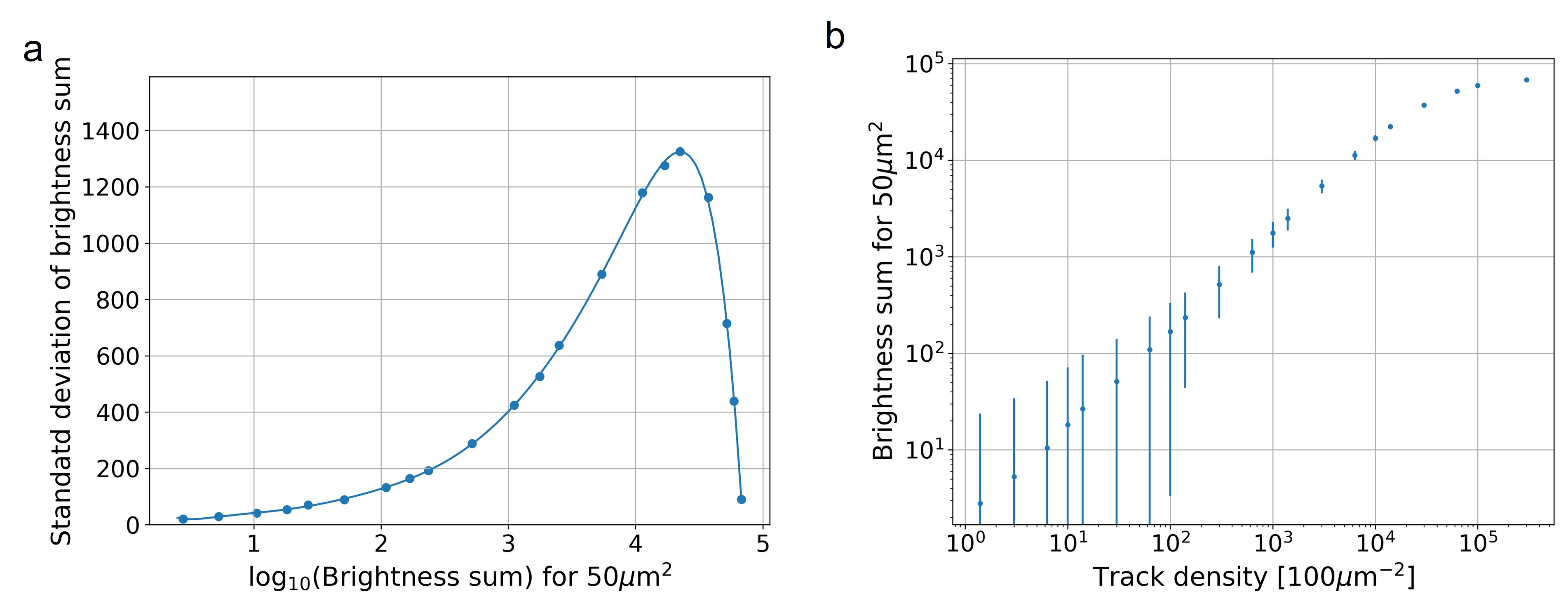}
    \caption{(a) Estimation of the standard deviation for the brightness sum values per 50 $\mu$m$^2$. The estimated standard deviation values were employed as length of the error bars for the brightness values shown in Figures \ref{fig:width_resolution_trap_fit}a, \ref{fig:width_resolution_trap_fit}b and \ref{fig:width_resolution_trap_fit}c. The curve shows the polynomial fitting function. (b) The brightness sum values per 50 $\mu$m$^2$ for varying track density conditions.}
    \label{fig:error_bar_estimation}
\end{figure}

The length of the statistical error bar of each data point is shown in Figures \ref{fig:width_resolution_trap_fit}a, \ref{fig:width_resolution_trap_fit}b and \ref{fig:width_resolution_trap_fit}c. The errors were estimated using the following procedure.
Data sets of simulated images were created under several track density conditions ranging from 1 track per 100 $\times$ 100 $\mu$m$^2$ to $3 \times$ $10^5$ tracks per 100 $\times$ 100 $\mu$m$^2$. Simulated images under track density of $1 \times$ $10^3$ and $1 \times$ $10^4$ tracks per 100 $\times$ 100 $\mu$m$^2$ are shown in Figures \ref{fig:errro_bar_ana_imgs}a and \ref{fig:errro_bar_ana_imgs}b, respectively. These simulated images were scaled down to 188 pixels $\times$ 188 pixels as discussed above. 
The mean and standard deviation values (shown in Figures \ref{fig:errro_bar_ana_projection}a and \ref{fig:errro_bar_ana_projection}b) were calculated from the brightness sum distribution of the resized images (shown in Figures \ref{fig:errro_bar_ana_imgs}a and \ref{fig:errro_bar_ana_imgs}b). Figure \ref{fig:error_bar_estimation}a shows the deduced standard deviation values as function of corresponding brightness values, while the blue curve shows the polynomial fitting. The deduced standard deviation values were employed as the length of error bar for the corresponding brightness values (shown in Figures \ref{fig:width_resolution_trap_fit}a, \ref{fig:width_resolution_trap_fit}b and \ref{fig:width_resolution_trap_fit}c). Figure \ref{fig:error_bar_estimation}b shows the brightness sum values as a function of track densities. The brightness sum was almost saturated in a high track density region around $10^5$ tracks per 100 $\times$ 100 $\mu$m$^2$.

\begin{figure}[htbp]
    \centering
    \includegraphics[width=0.9\linewidth]{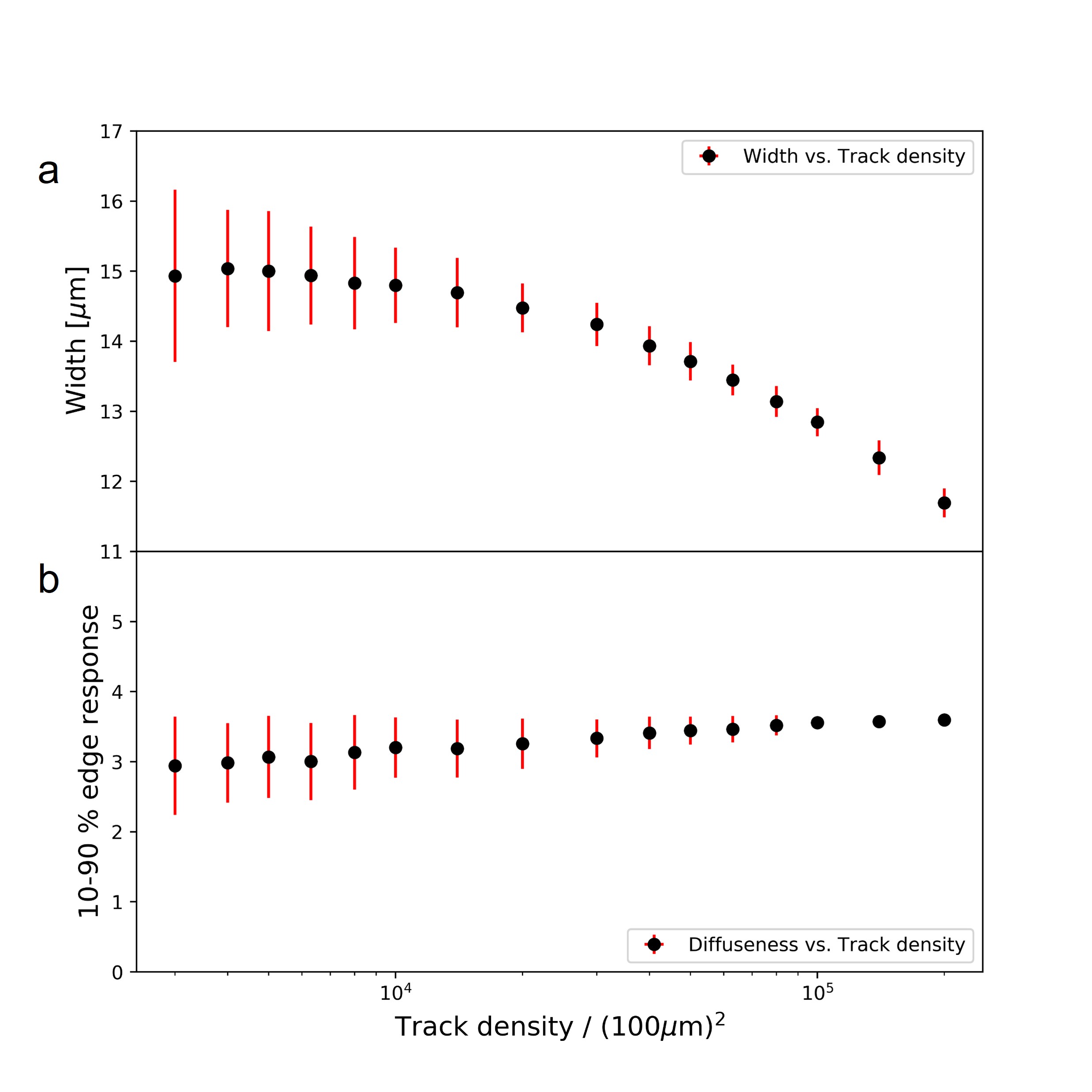}
    \caption{(a)  Width resolution to estimate track density for the neutron imaging application using the fine-grained nuclear emulsion, (b)  L(10-90\%) edge response for varying track density conditions.}
    \label{fig:width_resolution}
\end{figure}

The resolved width and L(10-90\%) edge response (displayed in Figure \ref{fig:trap_fit_example}) were used to evaluate the diffuseness as a function of track density as shown in Figures \ref{fig:width_resolution}a and \ref{fig:width_resolution}b, respectively. The length of error bar for each data point in these figures was estimated as standard deviation of corresponding brightness values. As the track density increases, the resolved width of the grating decreases due to finite length of the tracks in the FGNE. As shown in Figure \ref{fig:width_resolution}a, a track density of $10^4$ tracks per 100 $\times$ 100 $\mu$m$^2$ is sufficient for imaging applications using the current FGNE based NI system (also discussed in section \ref{FGNE_detector}). The L(10-90\%) edge response (shown in Figure \ref{fig:width_resolution}b) is relatively constant, and diffuseness of 3.33 $\pm$ 0.27 $\mu$m can be achieved under the accumulated track density of $3 \times 10^4$ tracks per 100 $\times$ 100 $\mu$m$^2$ for NI using the FGNE.

\begin{figure}[htbp]
    \centering
    \includegraphics[width=\linewidth]{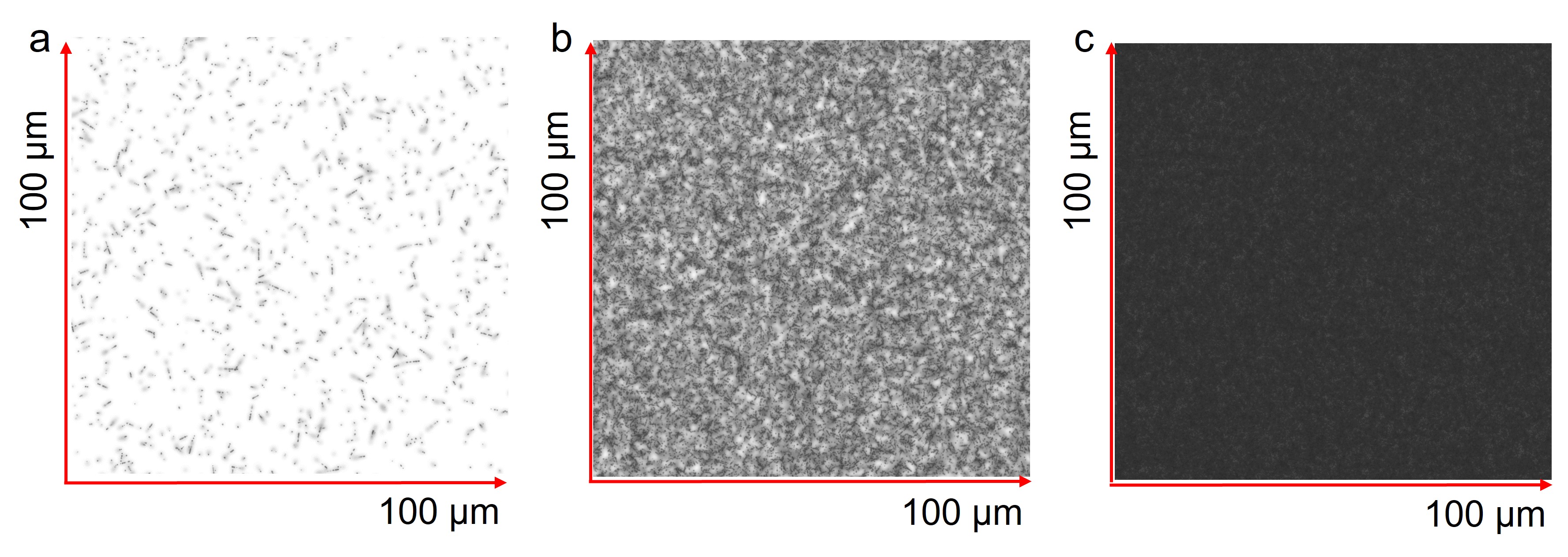}
    \caption{Simulated images for contrast resolution estimation under the accumulated track density of (a) $1 \times$ $10^3$, (b)  $3 \times$ $10^4$ and (c) $3 \times$ $10^5$ per 100 $\times$ 100 $\mu$m$^2$ for the neutron transmission rate of 95 \%.}
    \label{fig:constrast_resolution_sim_imgs}
\end{figure}

\begin{figure}[htbp]
    \centering
    \includegraphics[width=\linewidth]{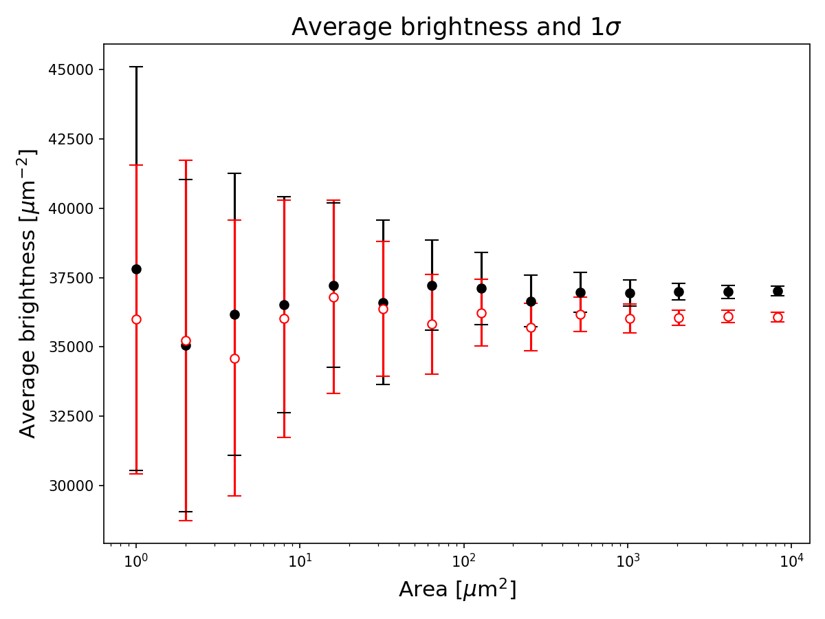}
    \caption{Contrast resolution for the image data set with a combination of 100 \% vs. 96 \% neutron transmission rate under track density of $3 \times 10^4$ per (100 $\mu$m)$^2$. The area which can be resolved is 2048 $\mu$m$^2$ as shown in the Table \ref{tab:Table_cons_resolution}. The black and red data points represent the average brightness for the neutron transmission rate of 100 \% and 96 \%, respectively.}
    \label{fig:contrast_resolution_check}
\end{figure}

\subsection{\label{estimation_contrast_resolution}Contrast resolution}

It is equally important to investigate the contrast resolution for the imaging applications using the FGNE. To investigate the contrast resolution, image data sets with track densities $1 \times 10^3$, $3 \times 10^3$, $1 \times 10^4$, $3 \times 10^4$, $1 \times 10^5$ and $3 \times 10^5$ per 100 $\times$ 100 $\mu$m$^2$ were generated with neutron transmission rates of 70 \%, 75 \%, 80 \%, 85 \%, 90 \%, 95 \%, 96 \%, 97 \%, 98 \% and a data set with 100 \% transmission. Figures \ref{fig:constrast_resolution_sim_imgs}a, \ref{fig:constrast_resolution_sim_imgs}b and \ref{fig:constrast_resolution_sim_imgs}c show the simulated images under the accumulated track density of $1 \times$ $10^3$, $3 \times$ $10^4$ and $3 \times$ $10^5$ per 100 $\times$ 100 $\mu$m$^2$, respectively, for the neutron transmission rate of 95 \%. The average brightness values in area ranging from  2$^0$, 2$^1$, .... , up to 2$^{13}$ $\mu$m$^2$, under the given neutron transmission, were compared with the corresponding average brightness values with the neutron transmission of 100 \%. Figures \ref{fig:contrast_resolution_check} show the comparison of the average brightness values for the combination of 100 \% vs. 96 \% neutron transmission rate under the accumulated track density of $3 \times 10^4$ per (100 $\mu$m)$^2$. The area capable of separation at least $1 \sigma$ per $\mu$m$^2$ was 2048 $\mu$m$^2$. The results of the contrast resolution for the combination of 100 \% vs. various neutron transmission rates under the accumulated track densities are summarized in Table \ref{tab:Table_cons_resolution}. Corresponding figures for the contrast resolution of each image data set, at the given neutron transmission rates and track densities, are shown in Appendix \ref{appendix}.

The simulation result for the image data sets with neutron transmission rate of 96 \% under the accumulated track density of $3 \times 10^4$ tracks per 100 $\times$ 100 $\mu$m$^2$ is shown in Table \ref{tab:Table_cons_resolution} and Figure \ref{fig:contrast_resolution_check}.
The simulation-based results can be compared with the results of NI of the gold wires in a crystal oscillator chip using the FGNE \cite{hirota2021}. Hirota \emph{et al.} visualized the gold wires with a diameter of 30 $\mu$m and length $\sim$ 1 mm under the accumulated track density of $3 \times 10^4$ tracks per 100 $\times$ 100 $\mu$m$^2$ and neutron absorption rate of 3.84 \%. For neutron transmission rate of 96 \%, the area which can be resolved is 2048 $\mu$m$^2$ as shown in Table \ref{tab:Table_cons_resolution} and Figure \ref{fig:contrast_resolution_check} and it is concluded that an object with an area much smaller than gold wire with a diameter of 30 $\mu$m and length $\sim$ 1 mm may be resolved using the FGNE.

\begin{table}
\caption{\label{tab:table1}Summary of the contrast resolution analysis and area capable of separation at least 1$\sigma$ per ($\mu$m$^2$) under the accumulated track densities per (100 $\mu$m)$^2$.}

\begin{tabular}{cccccccc}
\hline
&Trans. & $1 \times 10^3$ & $3 \times 10^3$ & $1 \times 10^4$ & 3 $\times$ $10^4$ & $1 \times 10^5$ & $3 \times 10^5$\\
\hline
100 \% vs. 
 & 70 \% & 1024 & 256 & 64 & 16 & 2 & 1 \\ 
 & 75 \%  & 1024  & 256 & 64 & 64 & 8 & 1 \\ 
 & 80 \%  & 2048  & 256 & 256 & 64 & 8 & 4\\ 
 & 85 \%  & 4096  & 1024 & 256 & 64 & 16 & 8 \\ 
 & 90 \%  & 8192  & 4096 & 512 & 512 & 64 & 32 \\ 
 & 95 \%  &   & 8192 & 4096 & 2048 & 512 & 128 \\ 
 & 96 \%  &   &  & 8192 & 2048 & 512 & 128 \\ 
 & 97 \%  &   &  &  & 4096 & 1024 & 256 \\ 
 & 98 \%  &   &  &  & 8192 & 2048 & 512 \\ 
\hline
\end{tabular}
\label{tab:Table_cons_resolution}
\end{table}

\section{\label{experiments}Experiments}
For the neutron irradiation, we used the low-divergence beam branch \cite{kenji2009} of BL05 in the Materials and Life Science Experimental Facility (MLF) at the Japan Proton Accelerator Research Complex (J-PARC) \cite{qubs1030009}. This is one of the most suitable beam branches in the MLF for this experiment because it provides a reasonably intense neutron beam owing to a coupled neutron moderator.

For NI of the gadolinium-based grating slit, the aperture of the slit  was 44 mm and 6 mm for the horizontal and vertical directions, respectively. The beam divergence in the horizontal and vertical directions was set to 0.3 mrad and 10 mrad, respectively. It is to be noted that the beam divergence parallel to the grating does not affect the blurring of the image. The neutron flux was $2\times10^6$ n/cm$^2$/s, and the neutron irradiation time was 2.8 h to accumulate approximately $2 \times 10^4$ tracks per (100 $\mu$m)$^2$. The distance between the gadolinium-based grating and the $^{10}$B${_4}$C layer was 1.5 mm. The detector and the grating were fixed by metal clips and mounted on a holder for optical elements on an optical table.

On the other hand for imaging applications using the FGNE, NI of the Siemens star test pattern was also performed. The aperture of the slit on the 12 m position was set to 4 mm and 6 mm for the horizontal and vertical directions, respectively, to avoid blurring due to vertical divergence. The corresponding horizontal and vertical divergences of the beam were 0.3 mrad and 1.0 mrad, respectively. The typical beam power and the neutron flux during the experiment were 0.7 MW and $10^5$ n/cm$^2$/s, respectively.
The irradiation time was 9 h to accumulate $\sim$  $1 \times 10^4$ tracks per (100 $\mu$m)$^2$.
The distance between the gadolinium layer of the Siemens star and the $^{10}$B${_4}$C layer in the detector was 1.1 mm. The NI experiments for the grating slit and the Siemens star test patter were carried out under a stable environmental temperature of 23.9 $\pm$ 0.1 $^\circ$C.


\subsection{\label{scanning}Scanning the emulsion in the neutron detector}
A dedicated optical microscope with an epi-illumination system was used to scan the emulsion layer on a non-transparent silicon substrate. The light source of this system was a 1 W LED with a wavelength of 455 nm. The objective lens was an oil-immersion type with a magnification of 100x with a numerical aperture (NA) of 1.45 (Nikon Plan Apo). The depth of field was derived as $2 \times \lambda$ / NA$^2 \simeq$ 0.43 $\mu$m. The focusing unit of this microscope system was controlled by a stepping motor. The microscope system acquired a set of sequential images along the optical axis with a pitch of 0.3 $\mu$m from the upper to the lower boundary of the emulsion layer with enough margin. The typical number of images to be taken  at a certain place was 63. To scan the specified volume of the emulsion layer in three dimensions, the system shifted the field of view in the horizontal direction at a 50 $\mu$m pitch by the two other stepping motors and repeated the image acquisition. These images were acquired as 8-bit gray scale images with a CMOS image sensor with 2048 pixels $\times$ 2048 pixels (SENTECH CMB401PCL) at speeds of up to 160 frames per second. The length of one side of the field of view was approximately 112 $\mu$m. Accordingly, a single-pixel corresponded to 0.055 $\mu$m. Optical aberration in the periphery of the field of view reduces the quality of the image. We, therefore, used the central region of the acquired gray scale image (i.e., 1024 pixels $\times$ 1024 pixels).

\section{\label{results}Results}
\subsection{\label{Gadolinium_grating}Gadolinium grating}
\begin{figure}[htbp]
    \centering
    \includegraphics[width=1.0\linewidth]{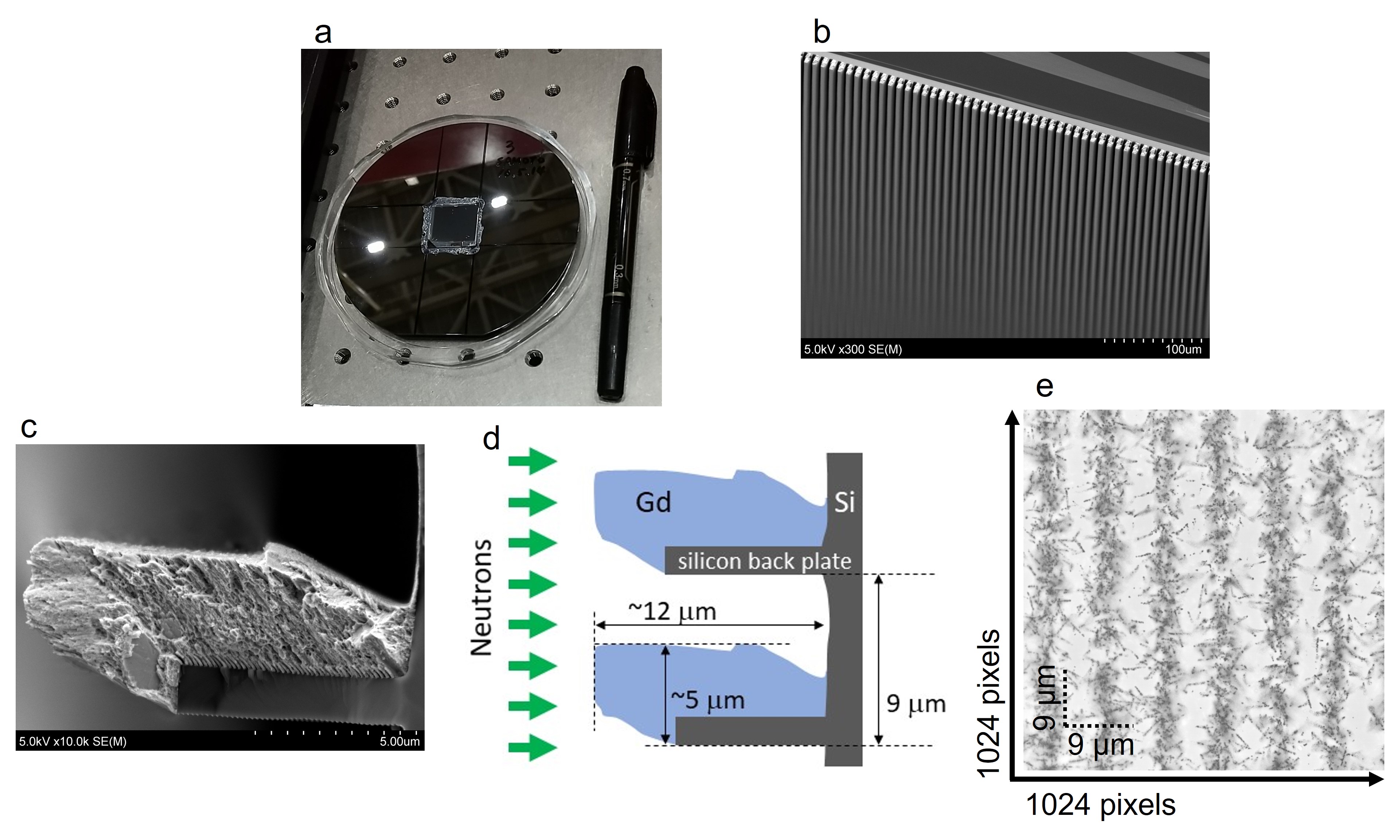}
    \caption{(a) Picture of a gadolinium-based grating with dimensions of 20 mm $\times$ 20 mm.  (b) Image  of the structure of the gadolinium teeth obtained by scanning electron microscopy (SEM). (c) SEM image of the cross-section of a single gadolinium tooth. (d) Schematic view of the gadolinium teeth.  (e) Optical micrograph (1024 pixels $\times$ 1024 pixels) of the neutron detector that recorded the tracks during neutron capture events through the grating.}
    \label{fig:gd_img2}
\end{figure}

The grating slit with a periodic structure of 9 $\mu$m was used for the NI using the FGNE, which was made for a Talbot--Lau interferometer for neutron phase imaging \cite{seki2017}. Figure \ref{fig:gd_img2}a shows the image of the gadolinium-based grating slit. It was formed on a 525 $\mu$m-thick silicon substrate and mounted at the centre of a 525 $\mu$m-thick five-inch silicon wafer. 
Figure \ref{fig:gd_img2}b is a scanning electron microscopy (SEM) image of a similar grating slit. A periodic structure of 9 $\mu$m (111 line pairs per mm) was formed on the silicon substrate. Figure \ref{fig:gd_img2}c is the SEM image of the cross-section of a gadolinium tooth of same type, while Figure \ref{fig:gd_img2}d shows a schematic view of the gadolinium teeth.   
Gadolinium teeth with a width of approximately 5 $\mu$m and empty spaces of 4 $\mu$m were formed periodically. The shape of  gadolinium tooth was not a perfect rectangular structure mainly because of the production process~\cite{Samoto_2019,SAMOTO201991}. Such grating can provide an inclusive assessment of the diffuseness including the effects of the resolution of the detector and imperfect shape of the gadolinium teeth.   
Each tooth was formed by depositing gadolinium from an oblique angle onto the ridge-like structures formed on the silicon substrate \cite{seki2017}. The direction of the irradiated neutron beam was from left to right. A portion of the incident neutron beam was absorbed by the gadolinium, while the remaining beam passed through the empty spaces. 
Figure \ref{fig:gd_img2}e shows a micrograph of tracks recorded in the emulsion layer. The dimensions and number of pixels of this image are 56 $\mu$m  $\times$ 56 $\mu$m and  1024 pixels $\times$ 1024 pixels, respectively. The periodic structure of gadolinium grating with 9 $\mu$m was clearly resolved and is visible.

\subsection{\label{img_processing}Image processing}
The acquired gray scale optical images of the grating slit were further processed.  The raw data obtained by the optical microscope were a set of 63 sequential images with 2048 pixels $\times$ 2048 pixels for each field of view. These optical images contained shades of dust on the image sensor and non-uniformity of the brightness due to optical conditions. In order to extract shades of dust, the
average brightness for each pixel in the 63 sequential images was calculated. To remove shades of dust and non-uniformity of the brightness, the pixel brightness value of each optical image was divided by the calculated average brightness value of the corresponding pixel. This image processing was applied to the  image data sets for 165 field of views. The images at the depth of the boundary between the sputtered carbon layer and the emulsion layer were selected for further analysis. The inner portion of 1024 pixels $\times$ 1024 pixels was cropped in these images. 

\subsection{\label{edge_response}10-90 \% edge response evaluation}
\begin{figure}[htbp]
    \centering
    \includegraphics[width=\linewidth]{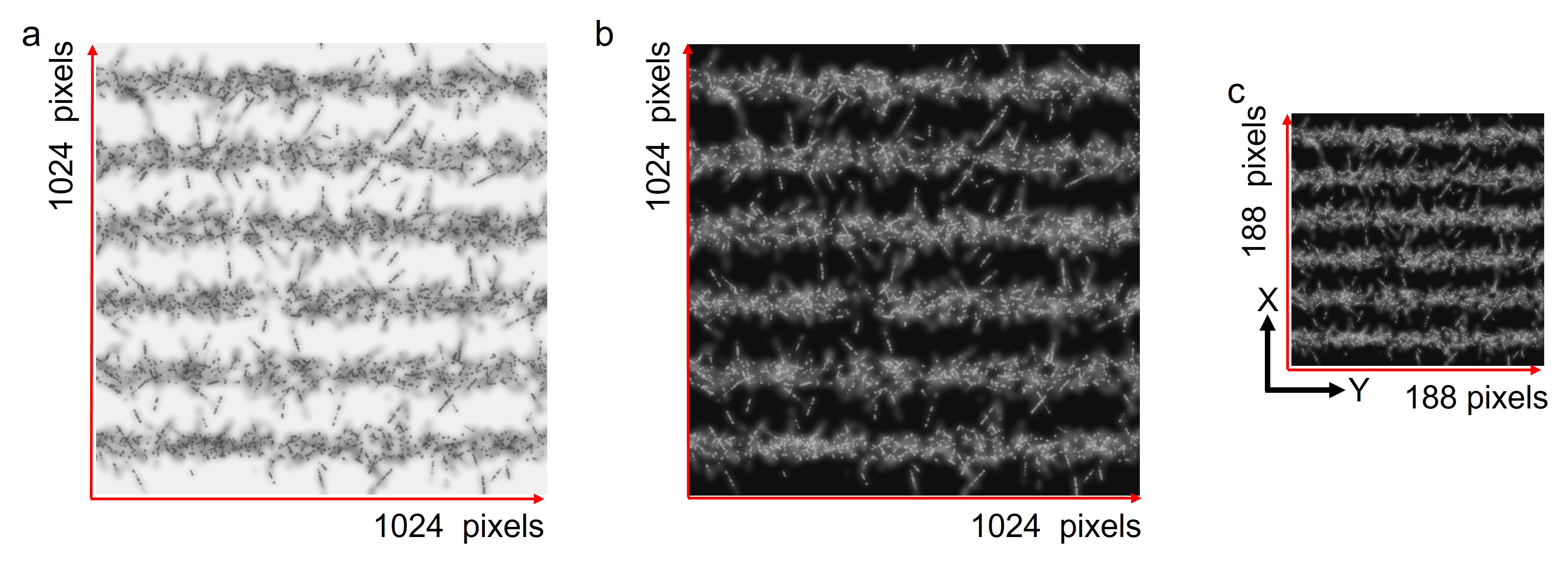}
    \caption{(a) Simulated image of the neutron detector that recorded the tracks during neutron capture events through the grating, (b) is the inverted image of (a) and (c) is the resized image of (b). The dimensions of this micrograph were 56 $\mu$m $\times$ 56 $\mu$m.}
    \label{fig:sim_imgs}
\end{figure}

\begin{figure*}[htbp]
    \centering
    \includegraphics[width=\linewidth]{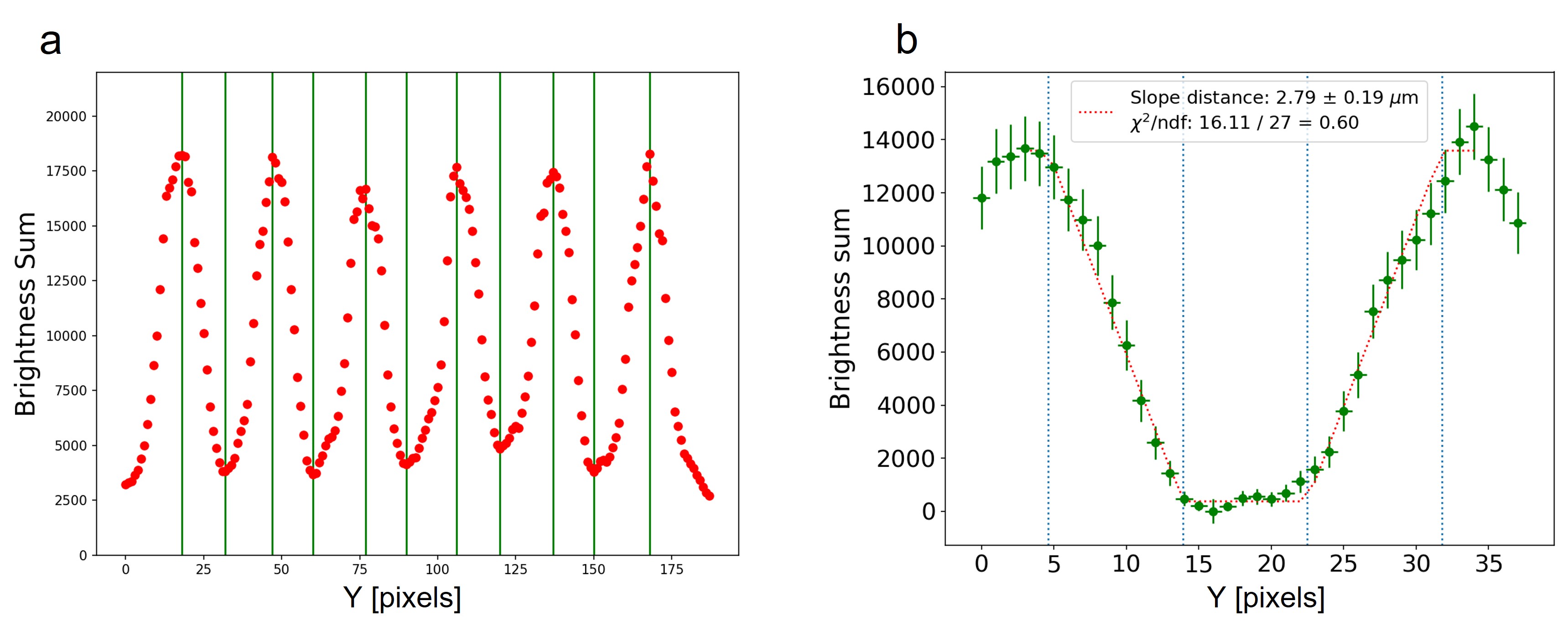}
    \caption{(a) Brightness sum in simulated image along the direction parallel to the grating (Y-direction) as a function of the X position. The green lines are the positions of the peaks and valleys, and the defined boundaries of  individual edge sections. (b) An example of data points of the edges and the trapezoid fitting. The red dotted lines represent the curves obtained by the trapezoid fitting.}
    \label{fig:sim_imgs_projection}
\end{figure*}

\begin{figure}[htbp]
    \centering
    \includegraphics[width=\linewidth]{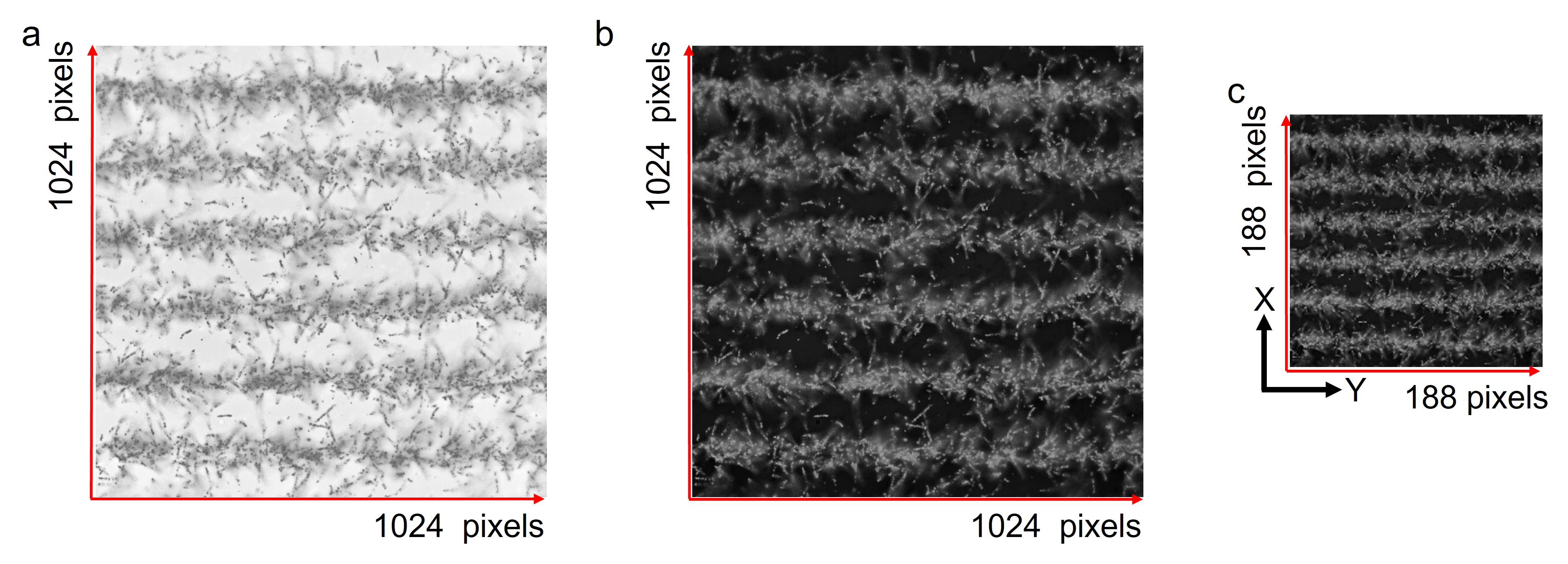}
    \caption{(a) Optical micrograph of the neutron detector that recorded the tracks during neutron capture events through the grating. (b) is the inverted image of (a), and (c) is the resized image of (b). The dimensions of this micrograph were 56 $\mu$m $\times$ 56 $\mu$m.}
    \label{fig:real_imgs}
\end{figure}

\begin{figure*}[htbp]
    \centering
    \includegraphics[width=\linewidth]{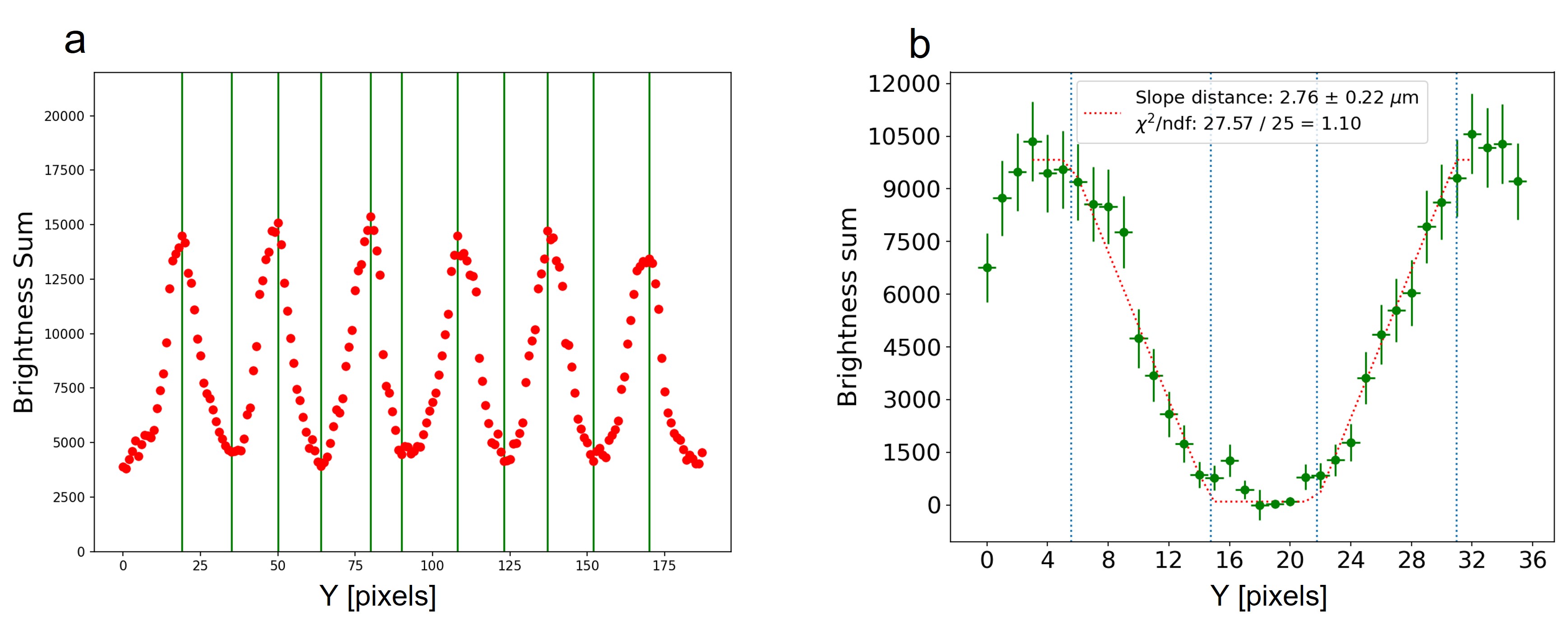}
    \caption{(a) Brightness sum in the optical micrograph along the direction parallel to the grating (Y-direction) as a function of the X position. The green lines are the positions of the peaks and valleys, and the defined boundaries of  individual edge sections. (b)  An example of data points of an edge profile and the trapezoid fitting. The red dotted lines represent the curves obtained by the trapezoid fitting.}
    \label{fig:real_imgs_projection}
\end{figure*}

\begin{figure}[htbp]
    \centering
    \includegraphics[width=\linewidth]{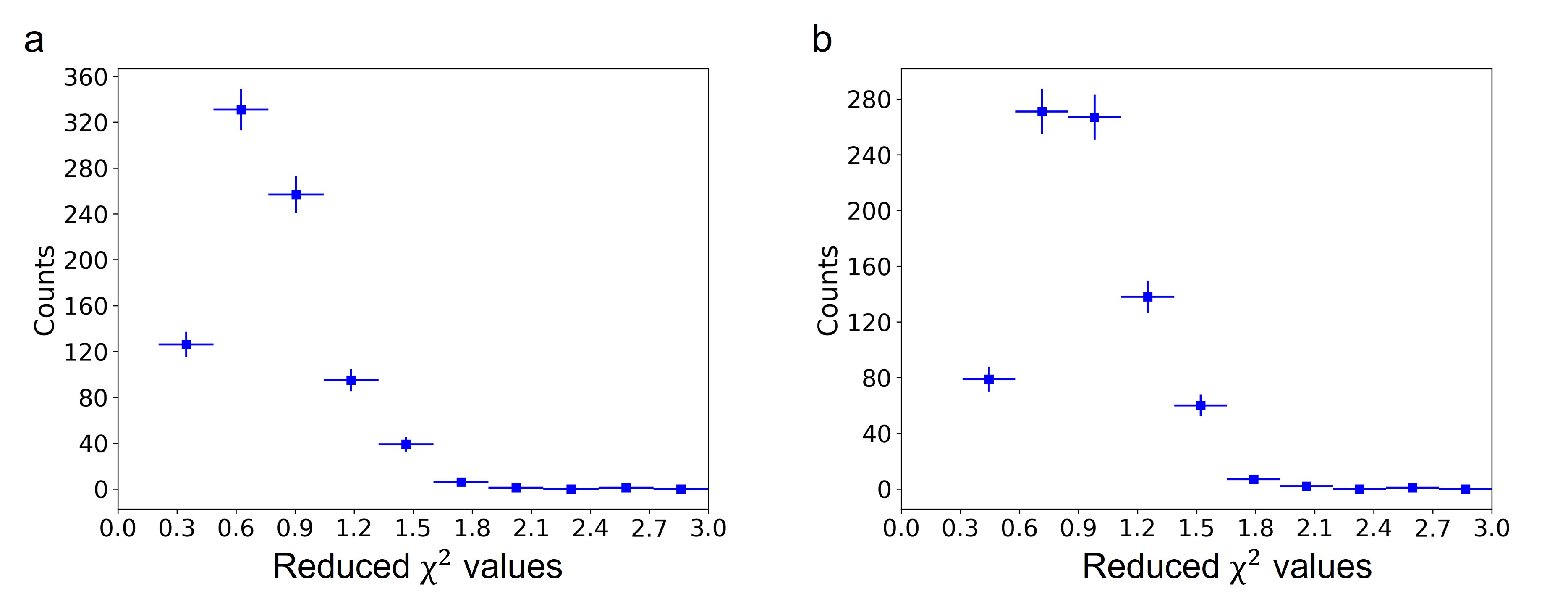}
    \caption{Distributions of the reduced $\chi^2$ values for the fitting with the trapezoid function (a) simulated images and (b) optical images. }
    \label{fig:red_chisq}
\end{figure}

\begin{figure*}[htbp]
    \centering
    \includegraphics[width=\linewidth]{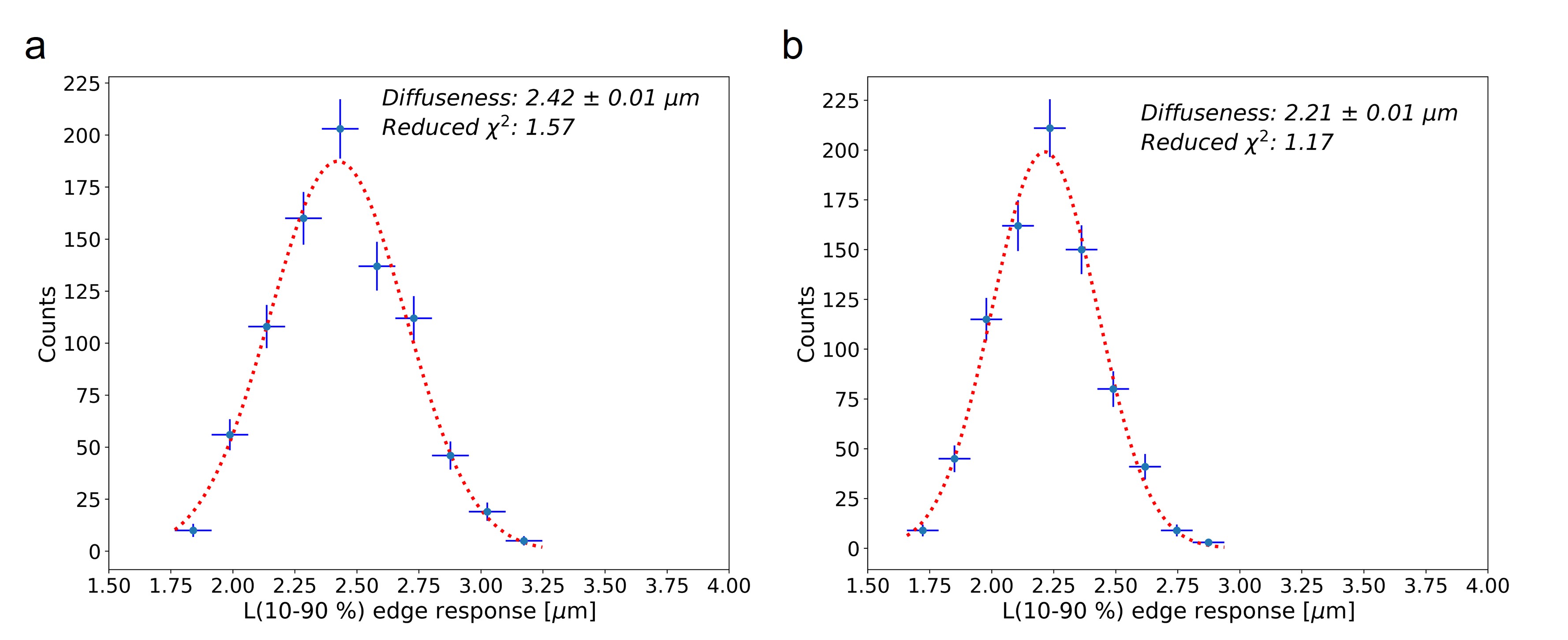}
    \caption{Distribution of the L(10-90\%) edge response is shown for (a) 165 optical images and (b) 165 simulated images. The red dotted lines in both panels are the Gaussian fitting curve.}
    \label{fig:edge_rise}
\end{figure*}

The L(10-90\%) edge response (shown previously  in  Figure \ref{fig:trap_fit_example}) of the grating slit with a periodic structure of 9 $\mu$m was evaluated using the simulated and optical images. The image processing, as explained in the previous subsection,  was employed to the optical images. Figures (\ref{fig:sim_imgs}a and \ref{fig:real_imgs}a) show the simulated and optical images of the grating slit, respectively. 
Figures (\ref{fig:sim_imgs}b and \ref{fig:real_imgs}b) show the inverted image of Figures (\ref{fig:sim_imgs}a and \ref{fig:real_imgs}a, respectively). The inverted images  were scaled down to 188 pixels $\times$ 188 pixels as discussed in the subsection \ref{width_resolution} and displayed in Figures (\ref{fig:sim_imgs}c and \ref{fig:real_imgs}c, respectively). Using these resized images, the brightness sum in the direction parallel to the grating (Y-direction) were calculated at specified X-value with a range of bin sizes. The green lines in Figures (\ref{fig:sim_imgs_projection}a and \ref{fig:real_imgs_projection}a) show the defined boundary positions of the peaks and valleys of each edge section in the simulated and real images, respectively. The data points between the two consecutive peaks were extracted and trapezoid fitting was employed in the subsequent analyses to calculate the L(10-90\%) edge response for evaluation of the diffuseness. For example, five edges were extracted from each image  shown in Figures (\ref{fig:sim_imgs_projection}a and \ref{fig:real_imgs_projection}a). 
Figures (\ref{fig:sim_imgs_projection}b and \ref{fig:real_imgs_projection}b) show the trapezoid fitting for the extracted data points. The red dotted lines in these figures show the fitting curves. The length of each error bar (shown in Figures \ref{fig:sim_imgs_projection}b and \ref{fig:real_imgs_projection}b) was estimated in a way similar to that explained in subsection \ref{width_resolution}. Figures (\ref{fig:red_chisq}a and \ref{fig:red_chisq}b) show the distributions of the reduced $\chi^2$ of the fitting for the simulated and optical images, respectively. Figures (\ref{fig:edge_rise}a and \ref{fig:edge_rise}b) show the distribution of L(10-90\%) edge diffuseness for the simulated and optical images, respectively. The 1 $\sigma$ of L(10-90\%) edge diffuseness was calculated for the inclusive assessment of the resolution. The red dotted lines in these images show the Gaussian fitting curve. The deduced values of the diffuseness distributions for simulated and optical images were 2.21 $\pm$ 0.01 and 2.42 $\pm$ 0.01 $\mu$m, respectively.  The L(10-90\%) edge response in 1 $\sigma$ for the simulated and optical images were 0.863 $\pm$ 0.004 $\mu$m and 0.945 $\pm$ 0.004 $\mu$m, respectively. The resolution of 0.945 $\pm$ 0.004 $\mu$m is an inclusive assessment that includes the effects of the resolution of the detector and the shape of the gadolinium tooth. Therefore this deduced resolution is the upper limit of the actual resolution, and might be better than the deduced resolution. Submicrometer spatial resolution maybe achieved using a fine-grained nuclear emulsion for neutron imaging because the deduced resolution is less than 1 $\mu$m.

As mentioned in subsection \ref{FGNE_detector},  the emulsion layer shrunk by a factor of 0.55 $\pm$ 0.1 after chemical development. With a shrinkage factor of 0.55, the deduced value of the diffuseness distribution for simulated images was 2.21 $\pm$ 0.01. In order to calculate the systematic error induced due to the shrinkage of the emulsion layer, additional simulated image data sets,  with a shrinkage factor of 0.4 and 0.6, were generated.  The deduced value of the diffuseness distribution was
$2.21 \pm 0.01 (stat.) \pm 0.02 (syst.)$.

The grating slit in the simulated images had a perfect rectangular structure. A fair agreement was observed between the diffuseness values of the simulated and optical images. As discussed in subsection \ref{Gadolinium_grating}, the grating slit did not posses a perfect rectangular structure. The slight difference in the diffuseness values could be attributed to the imperfectness in the shape of the grating slit.  The calculated diffuseness was in fair agreement for the simulated and optical images. Consequently the achievable diffuseness under the accumulated track density of $3 \times 10^4$ tracks per 100 $\times$ 100 $\mu$m$^2$ can be 3.33 $\pm$ 0.27 $\mu$m for NI of a simple bar like structure using the FGNE.

\section{\label{Siemens_star}Demonstration of imaging application using FGNE}
\begin{figure*}[htbp]
    \centering
    \includegraphics[width=1.0\linewidth]{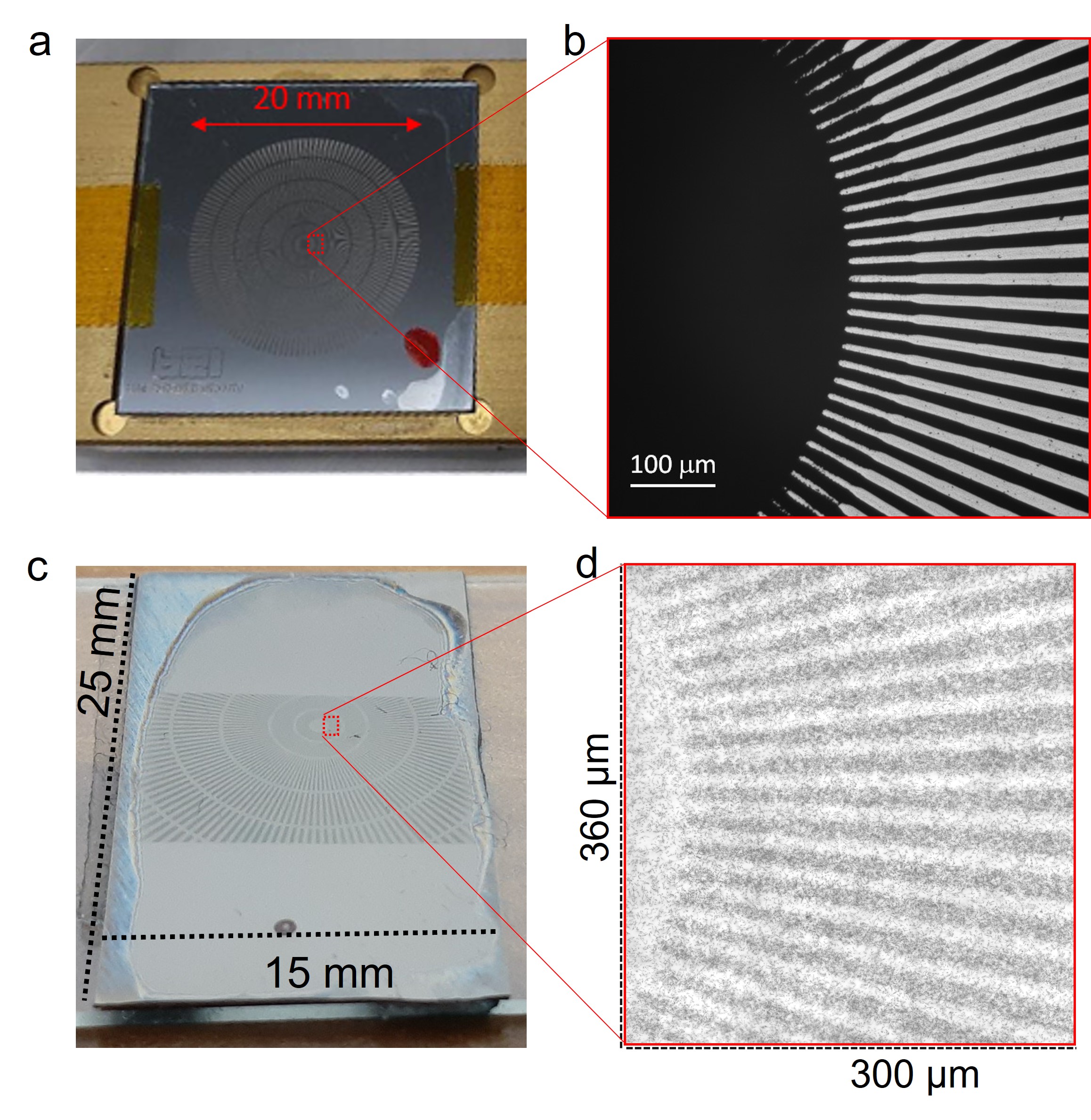}
    \caption{(a) Photograph of the Siemens star test pattern. It consists of a thin layer of gadolinium on a quartz substrate and is housed in an aluminum frame. (b) Optical image of the Siemens star test pattern. (c) Photograph of the developed detector irradiated with the neutron beam through the pattern. (d) Image of the recorded pattern in the nuclear emulsion showing the innermost edges. The sizes of the vertical and horizontal directions correspond to 360 and 300 $\mu$m, respectively. }
    \label{fig:PSI_patchwork_v01}
\end{figure*}

As discussed in the section \ref{experiments}, NI of the Siemens star test pattern was performed for the demonstration one of the applications of the proposed NI system.
Figure \ref{fig:PSI_patchwork_v01}a shows a photograph of the Siemens star test pattern used in our experiments.  
This pattern was designed by a group in the Paul Scherrer Institute (PSI), Villigen, Switzerland. A prototype of the test pattern is described in Ref. \cite{grunzweig2007}.
The pattern consisted of 128 spokes in a circular area of 20 mm diameter and was made of a thin layer ($\sim$ 5 $\mu$m) of gadolinium on a quartz wafer of thickness   0.7 mm.
The space between the spokes became narrower towards the center of the circle.
The innermost spoke had a period of 10 $\mu$m, which corresponds to 100 line pairs per millimeter. Figure \ref{fig:PSI_patchwork_v01}b shows an optical image of the Siemens star test pattern using optical microscope with a trans-illumination
system with a magnification of 20x, and a numerical aperture (NA) of 0.35.

The spatial resolution of an imaging device is assessed by the fine resolution of the intervals of the inner spokes. Figure \ref{fig:PSI_patchwork_v01}b shows a photograph of the developed detector irradiated with the neutron beam through the pattern. The pattern transferred on the emulsion layer is visible to the naked eye even before being observed under an optical microscope. The dark regions correspond to the sites where silver grains were developed by the emitted $^{4}$He or $^{7}$Li during the neutron absorption events. Other regions correspond to sites where the neutrons were absorbed by gadolinium on the pattern. The neutron absorption events in the $^{10}$B${_4}$C layer were suppressed. Figure \ref{fig:PSI_patchwork_v01}c shows an image of the nuclear emulsion, which recorded the thinnest spokes near the center of the pattern. As shown in Figure \ref{fig:PSI_patchwork_v01}b, the grating is not very well fabricated therefore the diffuseness for the edges of the grating was not evaluated quantitatively. One can note that the innermost spokes are clearly resolved as shown in Figure \ref{fig:PSI_patchwork_v01}d. Our NI results shows the ability of the FGNE to resolve the micrometer-scale structures with micrometer-scale resolution or even better. 

\section{\label{conclusion}Summary and outlook}
In this work, we presented the simulated results of neutron imaging using the fine-grained nuclear emulsion. It was concluded that a track density of $\sim$ $10^4$ tracks per 100 $\times$ 100 $\mu$m$^2$ was required for neutron imaging applications. The contrast resolution was investigated for various track densities under varying conditions of  neutron transmission rates. Our results estimated a contrast resolution and area capable of separation  1$\sigma$ per ($\mu$m$^2$) under various accumulated track densities per (100 $\mu$m)$^2$ and varying neutron transmission rates for the current neutron imaging system.  

The neutron imaging of gadolinium-based gratings with known geometries, such as a Siemens star test pattern and a grating slit with a periodic structure of 9 $\mu$m, were further reported in this work. These micrometer structured gratings were imaged and resolved successfully. 
The evaluated diffuseness for the edges of the grating slit under high accumulated track density conditions, for simulated and optical images, were 0.863 $\pm$ 0.004 $\mu$m and 0.945 $\pm$ 0.004 $\mu$m, respectively. The difference in the diffuseness values may be attributed to the shape of the gadolinium teeth. It is reminded that the grating slit is not a perfect rectangular structure because of the production process, as described in Ref. \cite{Samoto_2019,SAMOTO201991}.
The diffuseness values for simulated and optical images were in fair agreement and it may be concluded that the achievable resolution under the accumulated track density of $3 \times 10^4$ tracks per 100 $\times$ 100 $\mu$m$^2$ is 0.945 $\pm$ 0.004 $\mu$m. The obtained resolution for the optical images is an inclusive assessment that includes the effects of the resolution of the detector and the shape of the gadolinium tooth. It is concluded that the actual resolution may be better than the deduced resolution, and neutron imaging with submicrometer spatial resolution can be performed using the neutron detection system based on the fine-grained nuclear emulsion. In future, we plan to perform experiments for the exclusive assessment of simpler structure and evaluate the spatial resolution of the proposed neutron imaging system including neutron imaging of material objects.

\begin{acknowledgments}
The neutron experiment at the Materials and Life Science Experimental Facility of the J-PARC was performed under user programs (Proposal No. 2020B0223). The preliminary experiment on the detector was performed at the Research Reactor Institute, Kyoto University (Experiment No. R2138). J. Y. was supported by JSPS KAKENHI Grant Number JP18H05403 (Grant-in-Aid for Scientific Research on Innovative Areas 6005). The authors thank Takenao Shinohara of J-PARC and Yoshichika Seki of Tohoku University for lending the Siemens star test pattern and the grating, and Atsushi Momose and Tetsuo Samoto of Tohuku University for providing the SEM images shown in Figure \ref{fig:gd_img2} and for the discussions related to the grating. The authors thank Tatsuhiro Naka and Takashi Asada for providing us fine-grained nuclear emulsion gel, and for their advice regarding their use. The authors also thank Atsuhiro Umemoto and Ryuta Kobayashi from Nagoya University for their technical support in  using the optical microscope. The authors would like to thank Yukiko Kurakata of the High Energy Nuclear Physics Laboratory at RIKEN for providing administrative support for the entire project.

\end{acknowledgments}

\section*{Data Availability Statement}
The data that support the findings of this work are available from the corresponding author
upon reasonable request.

\begin{center}
\renewcommand\arraystretch{1.2}

\end{center}

\bibliography{aipsamp}

\newpage
\appendix

\section{\label{appendix}Appendix}
\begin{figure*}
    \centering
    \includegraphics[width=\linewidth]{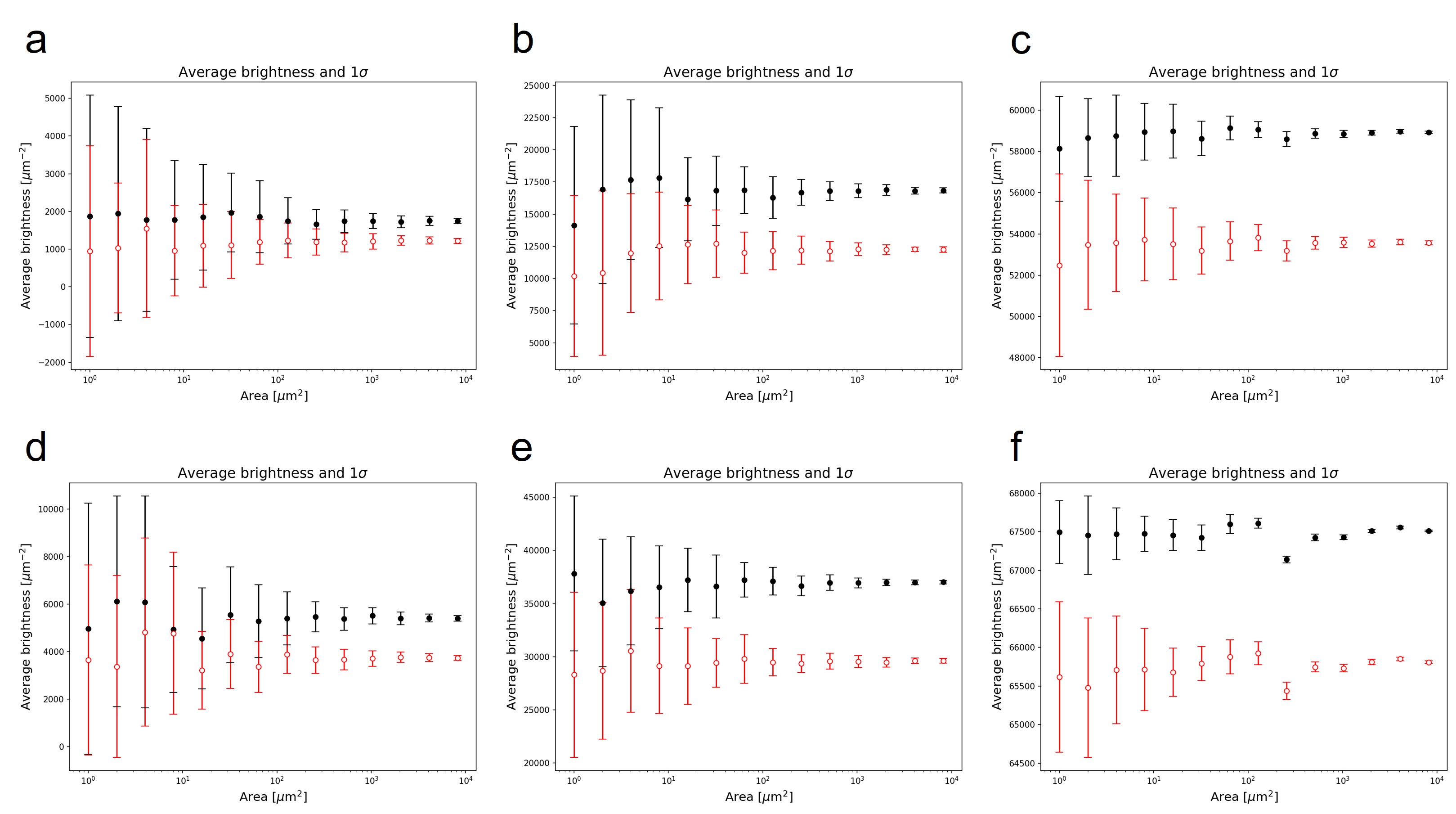}
    \caption{Contrast resolution for the image data set for the neutron transmission rate with a combination of 100 \% vs. 70 \% under the track density of (a) $1 \times 10^3$ (b) $3 \times 10^3$) (c) $1 \times 10^4$) (d) $3 \times 10^4$ (e) $1 \times 10^5$ (f) $3 \times 10^5$ per (100 $\mu$m)$^2$. The areas capable of separation at least 1$\sigma$ per ($\mu$m$^2$) under the accumulated track densities per (100 $\mu$m)$^2$ are shown in Table \ref{tab:Table_cons_resolution}.}
    \label{fig:contrast_resolution_trans70}
\end{figure*}

\begin{figure*}[htbp]
    \centering
    \includegraphics[width=\linewidth]{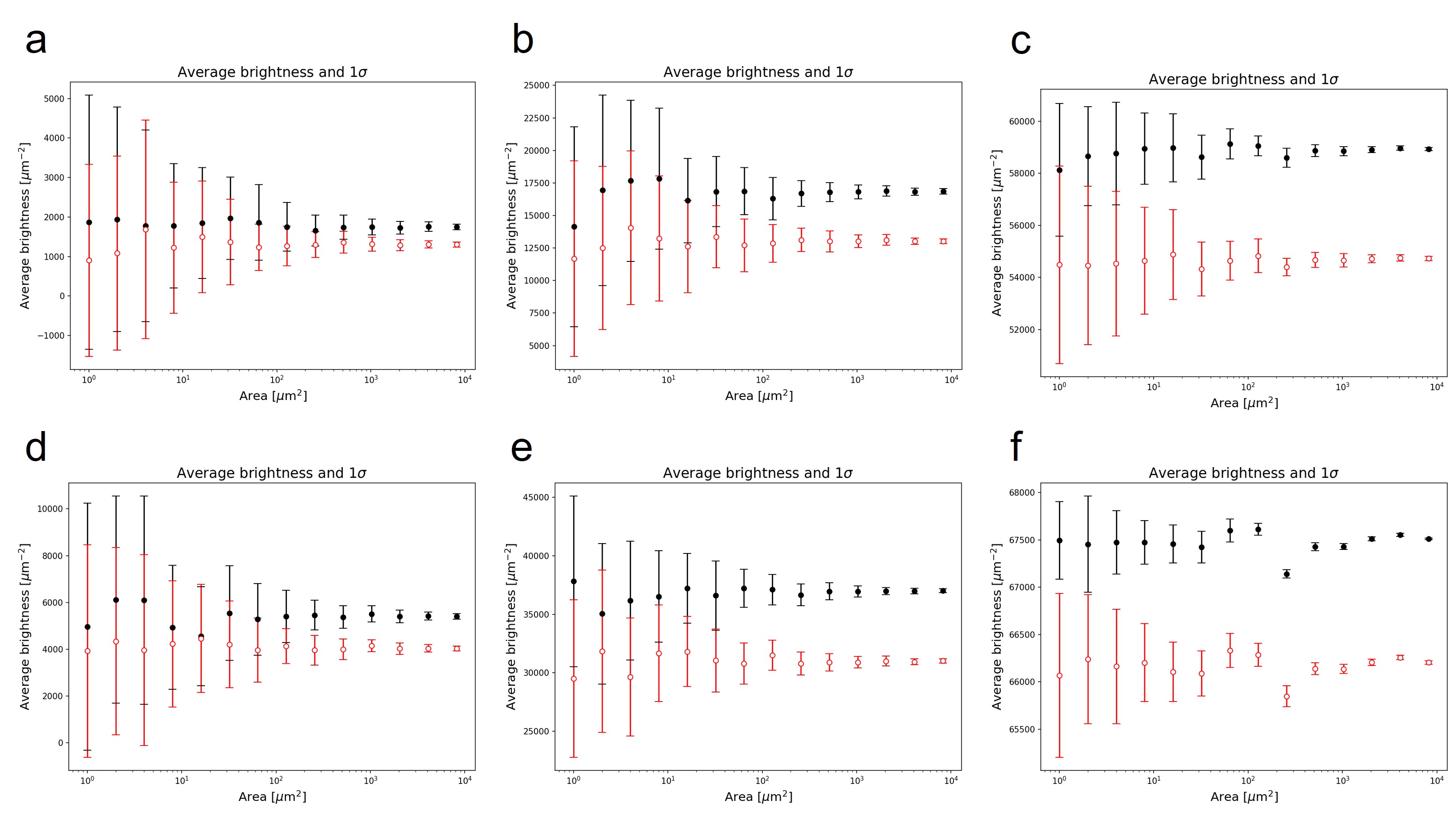}
    \caption{Same as  Figure \ref{fig:contrast_resolution_trans70} under the neutron transmission rate of 75 \%.}
    \label{fig:contrast_resolution_trans75}
\end{figure*}
\begin{figure*}[htbp]
    \centering
    \includegraphics[width=\linewidth]{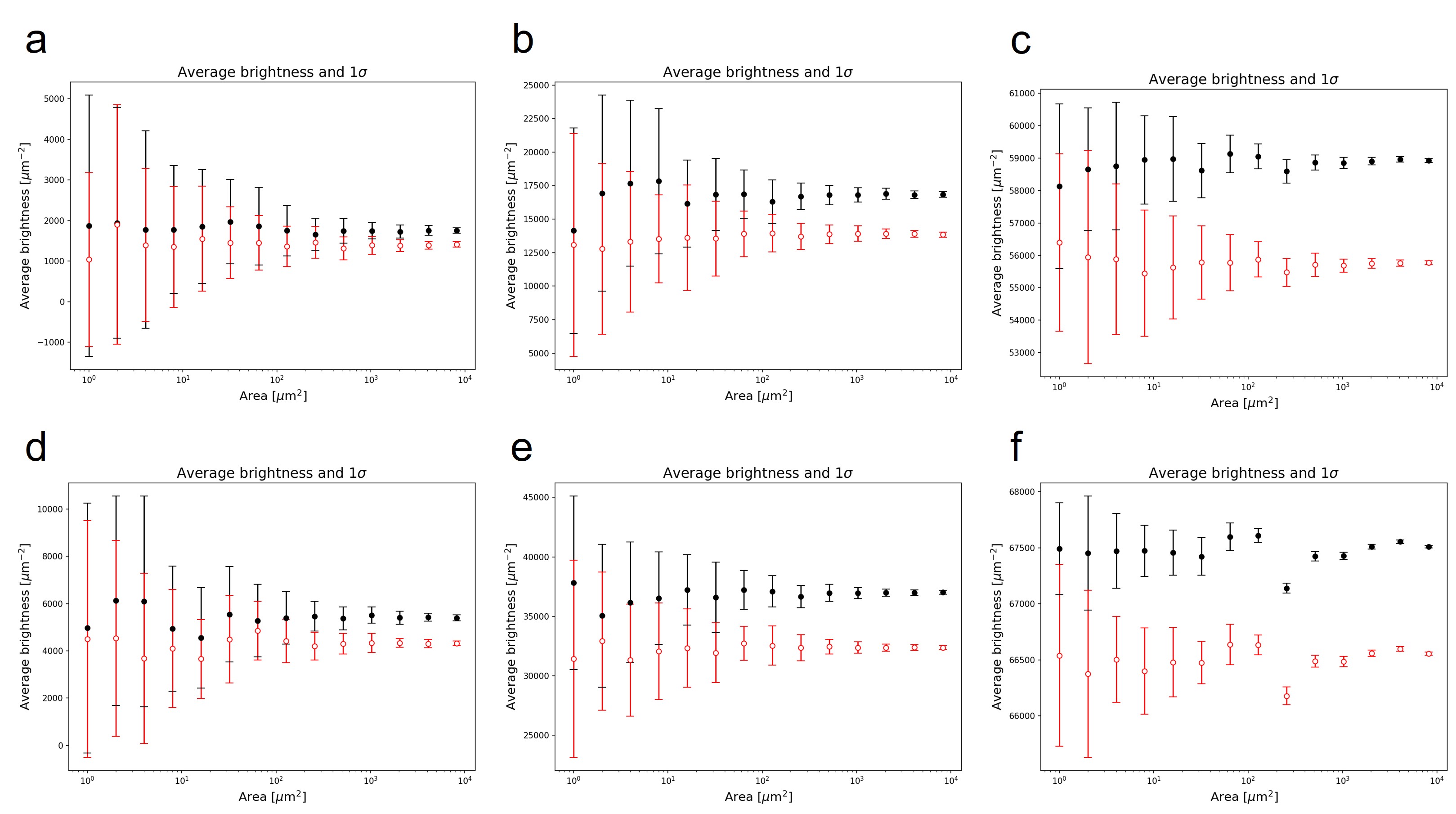}
    \caption{Same as Figure \ref{fig:contrast_resolution_trans70} under the neutron transmission rate of 80 \%.}
    \label{fig:contrast_resolution_trans80}
\end{figure*}

\begin{figure*}[htbp]
    \centering
    \includegraphics[width=\linewidth]{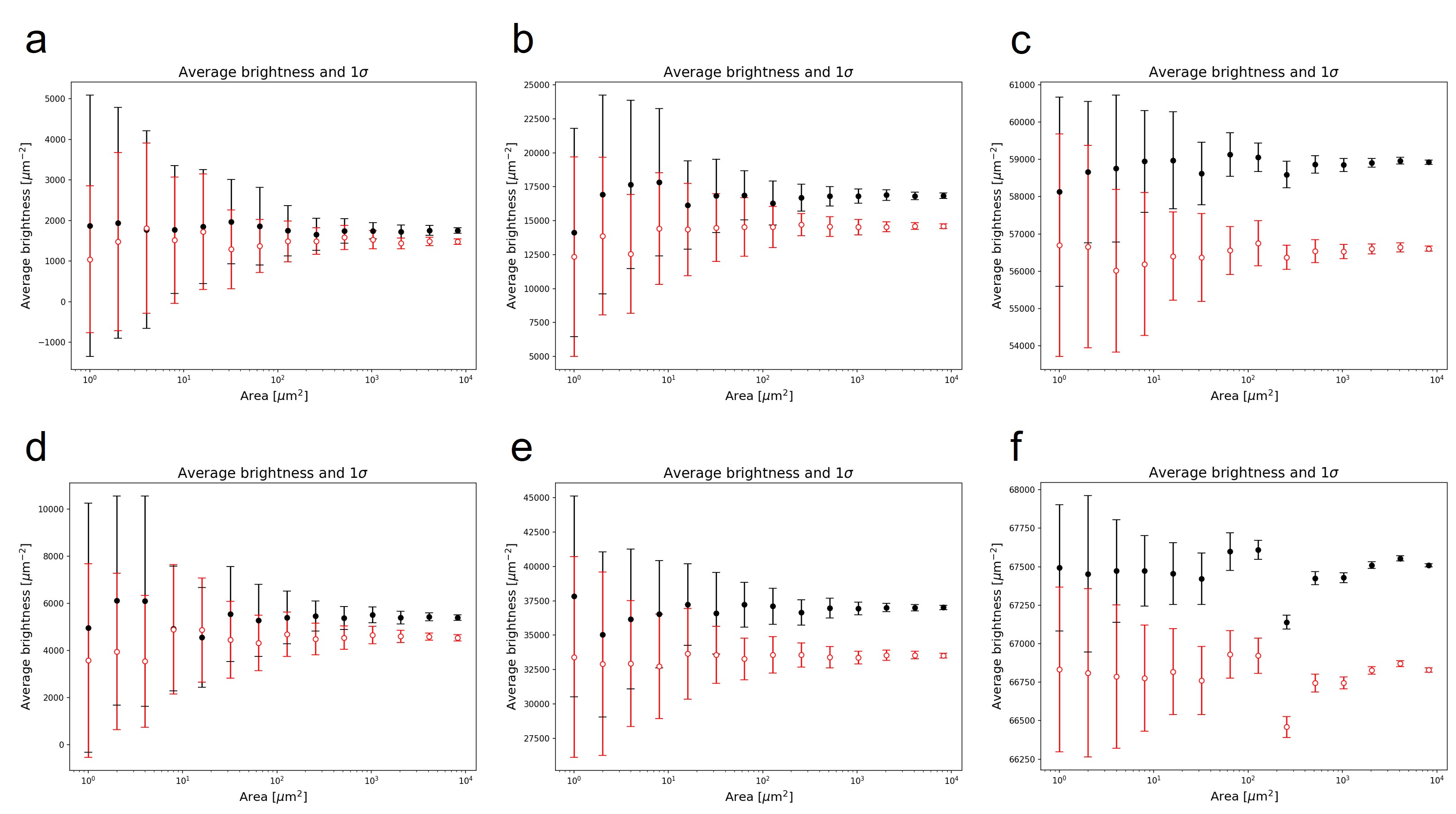}
    \caption{Same as Figure \ref{fig:contrast_resolution_trans70} under the neutron transmission rate of 85 \%.}
    \label{fig:contrast_resolution_trans85}
\end{figure*}

\begin{figure*}[htbp]
    \centering
    \includegraphics[width=\linewidth]{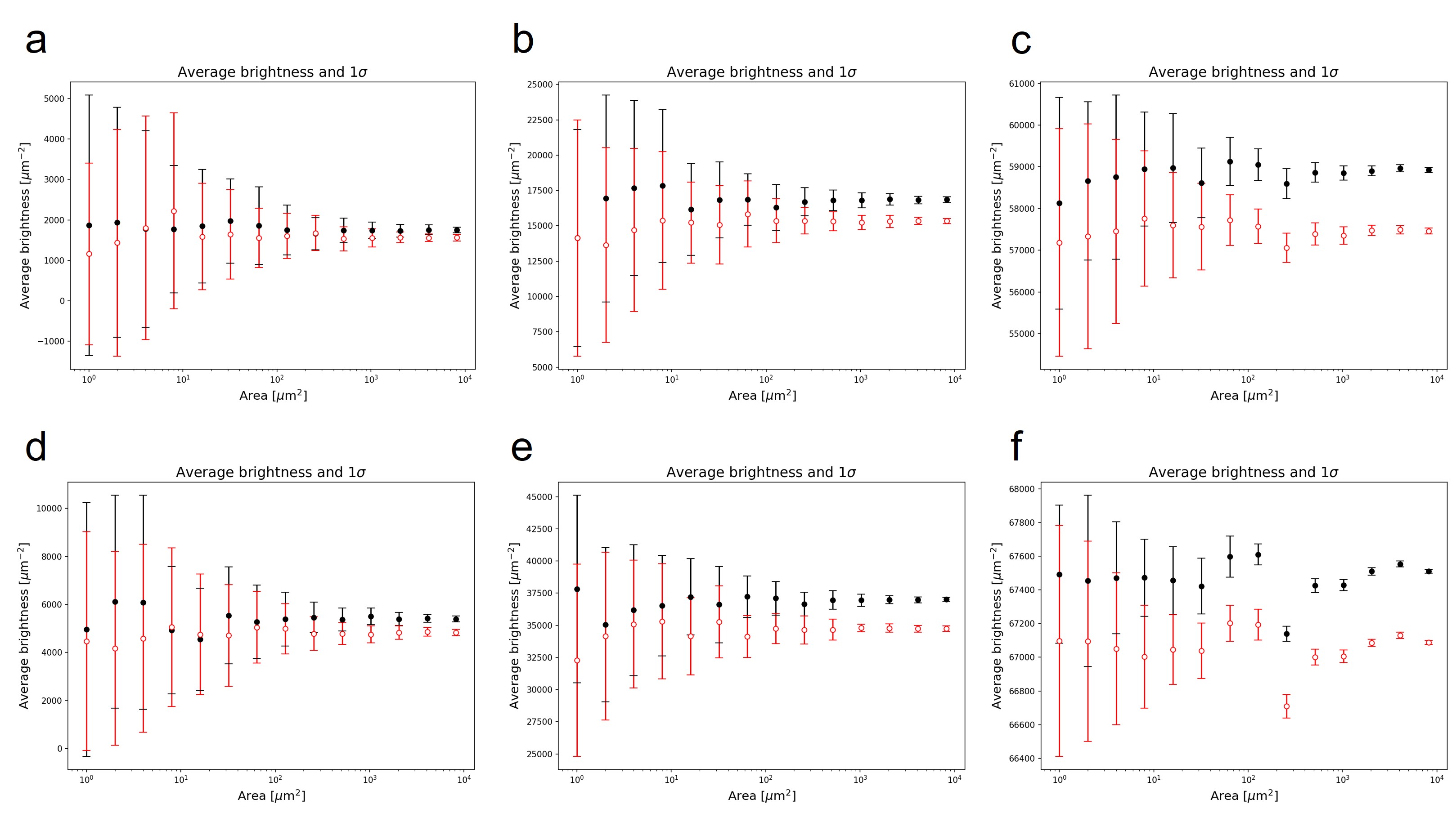}
    \caption{Same as Figure \ref{fig:contrast_resolution_trans70} under the neutron transmission rate of 90 \%.}
    \label{fig:contrast_resolution_trans90}
\end{figure*}

\begin{figure*}[htbp]
    \centering
    \includegraphics[width=\linewidth]{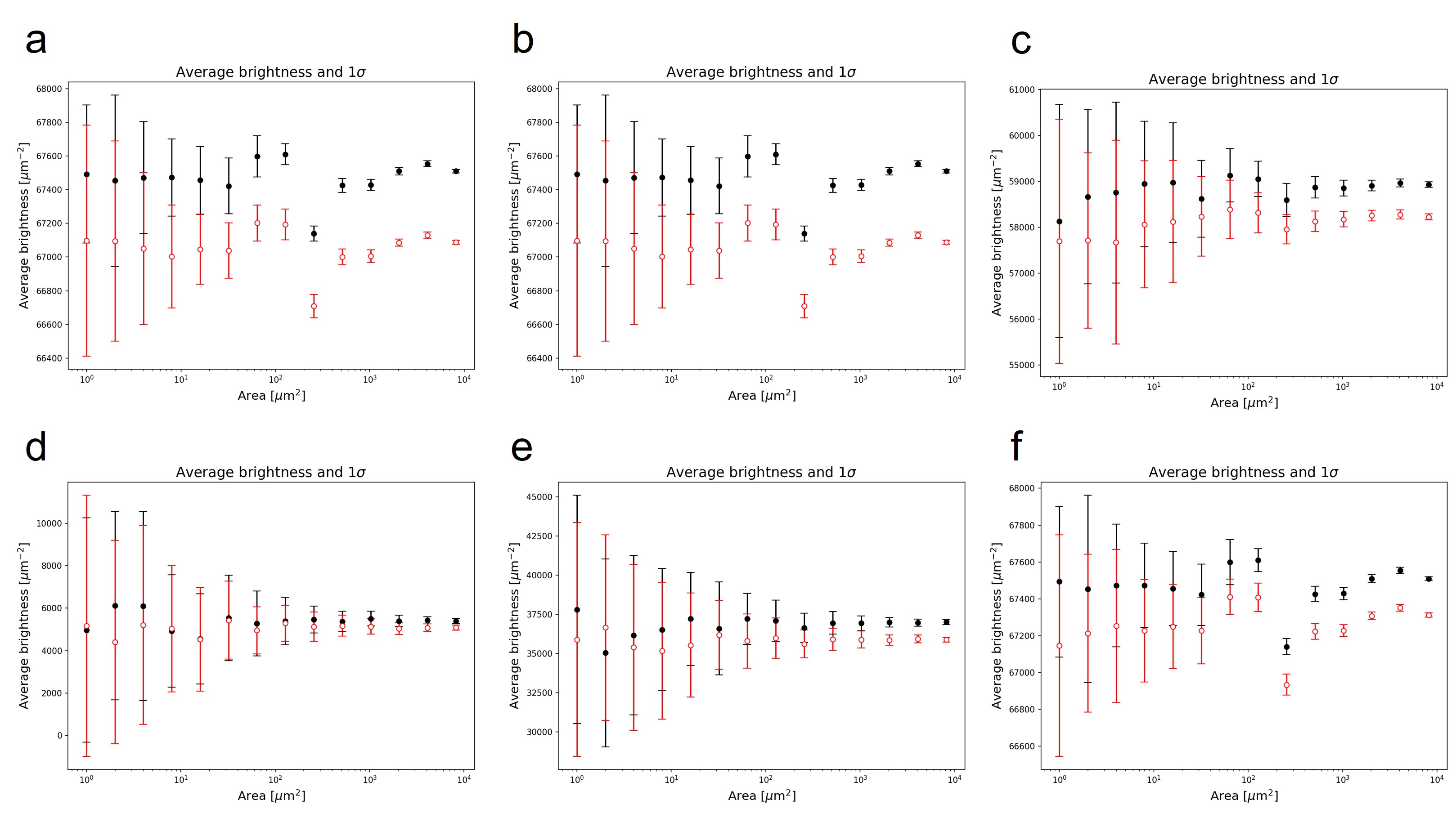}
    \caption{Same as Figure \ref{fig:contrast_resolution_trans70} under the neutron transmission rate of 95 \%.}
    \label{fig:contrast_resolution_trans95}
\end{figure*}

\begin{figure*}[htbp]
    \centering
    \includegraphics[width=\linewidth]{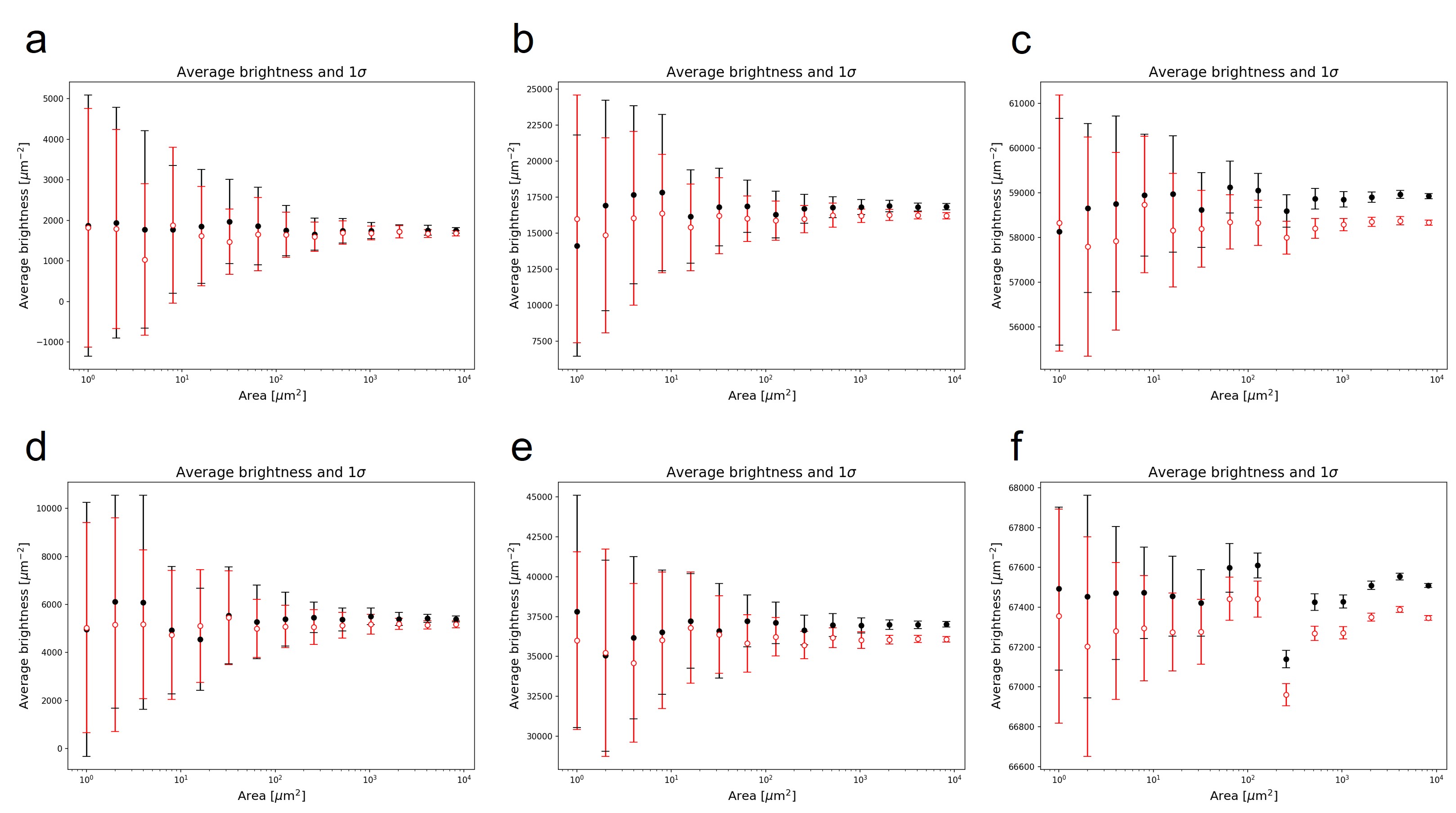}
    \caption{Same as Figure \ref{fig:contrast_resolution_trans70} under the neutron transmission rate of 96 \%.}
    \label{fig:contrast_resolution_trans96}
\end{figure*}
\begin{figure*}[htbp]
    \centering
    \includegraphics[width=\linewidth]{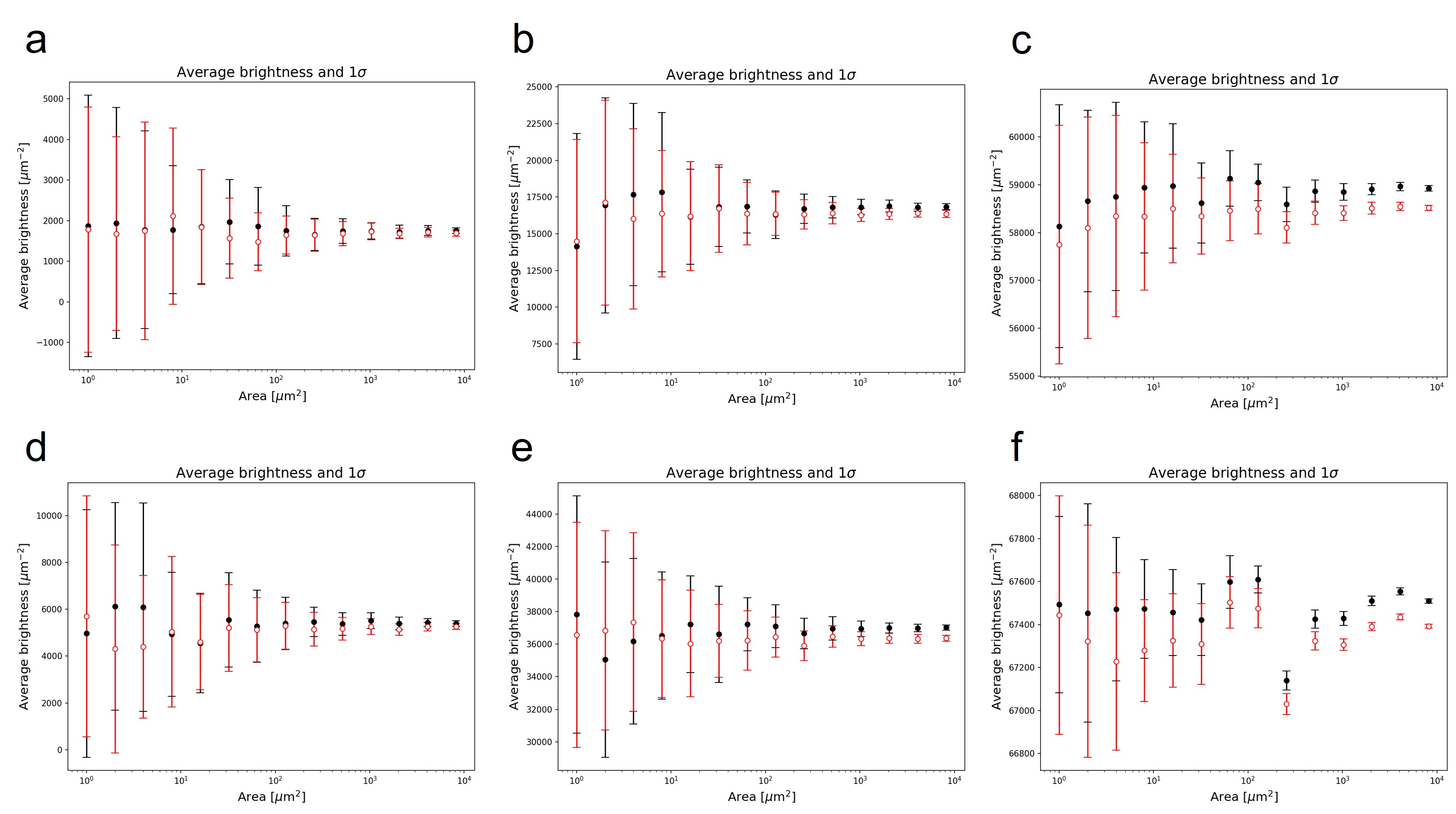}
    \caption{Same as Figure \ref{fig:contrast_resolution_trans70} under the neutron transmission rate of 97 \%.}
    \label{fig:contrast_resolution_trans97}
\end{figure*}

\begin{figure*}[htbp]
    \centering
    \includegraphics[width=\linewidth]{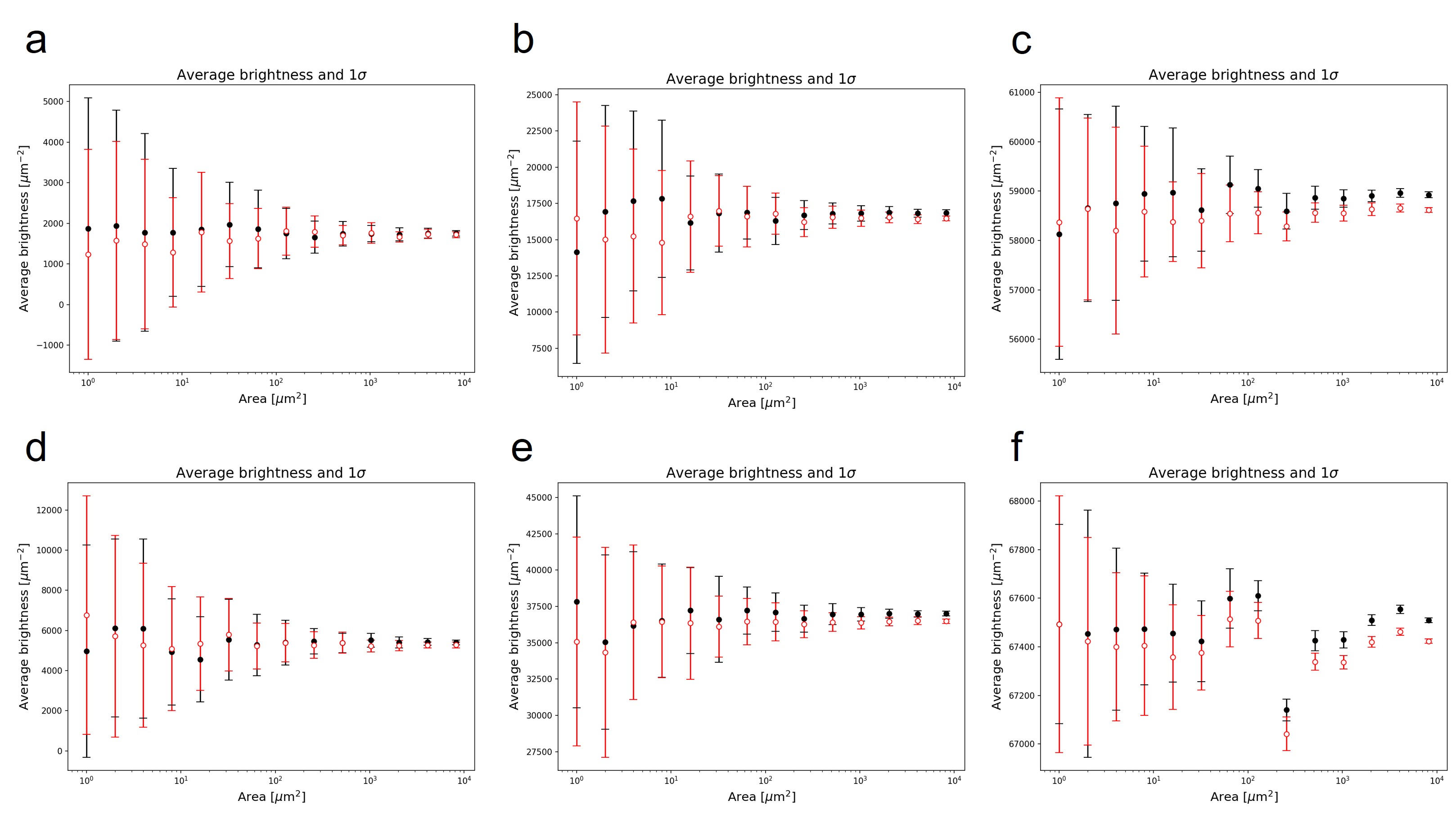}
    \caption{Same as Figure \ref{fig:contrast_resolution_trans70} under the neutron transmission rate of 98 \%.}
    \label{fig:contrast_resolution_trans98}
\end{figure*}

\end{document}